\documentclass{article}
\usepackage[utf8]{inputenc}
\usepackage[preprint]{gpsmdms_2022}
\usepackage{lipsum}
\usepackage{graphicx}
\usepackage{amsfonts}
\usepackage{oubraces}
\usepackage[numbers]{natbib}
\usepackage[hyphens]{url}
\usepackage[hidelinks]{hyperref}
\usepackage{xcolor, soul}
\hypersetup{breaklinks=true}
\usepackage{setspace}
\usepackage{amsmath}
\usepackage{caption}
\usepackage{subcaption}
\usepackage{pifont}
\usepackage[ruled, lined, linesnumbered, commentsnumbered, longend]{algorithm2e}
\title{Multi-Fidelity Data-Driven Design and Analysis of Reactor and Tube Simulations}

\author{Tom Savage\\
Department of Chemical Engineering\\
Imperial College London\\
London,UK\\
\texttt{trs20@ic.ac.uk} 
\And 
Nausheen Basha\\
Department of Chemical Engineering\\
Imperial College London\\
London,UK\\
\texttt{nausheen.basha@imperial.ac.uk}\\
\And
Jonathan McDonough\\
School of Engineering\\
Newcastle University\\
Newcastle,UK\\
\texttt{jonathan.mcdonough@ncl.ac.uk}
\And
Omar K Matar\\
Department of Chemical Engineering\\
Imperial College London\\
London,UK\\
\texttt{o.matar@imperial.ac.uk}
\And
Ehecatl Antonio del Rio Chanona\footnote{Corresponding Author}\\
Department of Chemical Engineering\\
Imperial College London\\
London,UK\\
\texttt{a.del-rio-chanona@imperial.ac.uk}\\
}

\begin{document}

\maketitle
\begin{abstract}
Optimizing complex reactor geometries is vital to promote enhanced efficiency.
We present a framework to solve this nonlinear, computationally expensive, and derivative-free problem.
Gaussian processes are used to learn a multi-fidelity model of reactor simulations correlating multiple continuous mesh fidelities. 
The search space of reactor geometries is explored through lower fidelity simulations, evaluated based on a weighted acquisition function, trading off information gain with cost.
Within our framework, DARTS, we derive a novel criteria for dictating optimization termination, ensuring a high fidelity solution is returned before budget is exhausted. 
We investigate the design of helical-tube reactors under pulsed-flow conditions, which have demonstrated outstanding mixing characteristics.
To validate our results, we 3D print and experimentally validate the optimal reactor geometry, confirming mixing performance. 
Our framework is applicable to a wide variety of expensive simulation-based optimization problems, supporting the design of the next generation of highly parameterized chemical reactors.

\end{abstract}

\section{Introduction}

Processes to transform raw materials into valuable products will always need to occur; as markets grow, practitioners seek to identify novel economically and environmentally friendly alternatives to existing solutions.
At the core of these processes, chemical reactors have long been studied and optimized for industrial chemical synthesis.
However, as batch processes are being made continuous \citep{Sarkis2021,Papathanasiou2019}, and specialty products such as biologically-derived chemicals enter industrial viability \citep{Tiwari2018}, there is a need for the design of novel case-specific reactors.

Novel reactor configurations have increasingly been considered for chemical synthesis, for example, microfluidic reactors \citep{Elvira2013,Losey2002}, and mesoscale reactors \citep{Santana2020}.
Microfluidic reactors can enable finer control over local conditions resulting in increased product selectivity, and improved heat transfer resulting in more sustainable processes \citep{Elvira2013}.
3D printed mesoscale reactors have been proposed as next-generation alternatives to traditionally manufactured designs, lending to their large potential design space.
3D printed reactors for biodiesel production have been shown to provide high yields \citep{Santana2020}, and helical coil-in-coil reactors, unable to be manufactured otherwise, have been shown to demonstrate `excellent plug-flow behavior' \citep{McDonough2019a,McDonough2019b,McDonough2019coilincoil}.

The modeling and design of traditional chemical reactors has largely been considered an art \citep{Green2007-vu}, with small improvements in performance resulting in wider impacts on product yield, sustainability, and economic costs.
However, with the promise of new reactors comes the need for new analytical techniques to model and optimize in increasingly complex design spaces. 

Chemical reactors have been investigated through computational fluid dynamics (CFD) simulations, where systems of partial differential equations (PDEs) with large degrees of freedom are solved iteratively, resulting in large computational costs. 
In addition to being expensive, gradient information is practically unavailable. 
In some simulation-based scenarios, optimization quantities can be obtained directly from a simulation \citep{LIEBECK1970}. 
As such, the adjoint method can be applied to derive gradient quantities which can be used within a gradient-based local optimization scheme such as the Newton method or BFGS \citep{Kenway2019,Mller2005,nocedal2006numerical}. 
However, in chemical reactor-based domains, performance is often quantified through a secondary quantity. 
For example, a tracer is simulated to pass through the reactor, and a value\footnote{For example the number of equivalent tanks-in-series} is then subsequently derived from this concentration profile through an additional fitting procedure \citep{savage2022}. 
This results in a scenario in which the gradient of a simulation is practically inaccessible, and derivative-free optimization must be applied.

Derivative-free optimization has found significant application in domains where mathematical expressions or gradients are unavailable. 
Examples include the optimization of proprietary chemical process software \citep{Savage2021,caballero2008algorithm}, chemical reaction optimization \citep{Felton2021}, real time optimization \citep{vandeBerg2022,Chanona2021}, and topology optimization of two-dimensional chemical reactor channels \citep{Cai2021}.
With the advent of new technologies in reactor design, reactor geometries are becoming highly-parameterized, resulting in higher-dimensional, more complex derivative-free optimization problems. 
As such, there exists significant scope for a robust, domain-specific approach for the optimization of simulated chemical reactors to support the next generation of sustainable chemical processes.

In many real-world and simulated engineering systems, differing quality evaluations of quantities of interest exist. 
Reactor performance can be quantified by a correlation of dimensionless numbers, a CFD simulation, or a pilot-scale experiment. 
These all attempt to capture the true underlying performance of a system, with differing accuracies and associated costs. 
Taking the view that only the industrial-scale reactor in its intended setting will provide a true evaluation of performance: any approximation to this, including pilot and lab-scale experiments, simulations, and basic calculations all become valid lower-fidelity evaluations which may be used simultaneously for design and optimization.
For CFD simulations of a reactor, fidelities are most often associated with the number of finite element cells in a simulation as they dictate the accuracy and computational cost.
By motivating the notion that all predicted, unmeasured quantities derive from a lower-fidelity approximation to a desired high-fidelity function, it becomes pertinent to investigate methodologies that apply these approximations to learn about the true system of interest.

The purpose of this article is to formulate the design of a simulated helical-tube reactor as a multi-fidelity black-box optimization problem.
We present a novel approach building upon previous work and demonstrate our framework for the Design and Analysis of Reactor and Tube Simulations, \textsc{DARTS}, by applying it for the simultaneous optimization of a helical-tube reactor geometry and operating conditions. 
In motivating our methodology, we derive a number of new criteria for monitoring the progress and dictating the termination of multi-fidelity Bayesian optimization. 
Our approach is extensible to a large number of simulation-based design problems, and our industrially relevant application is among the largest presented in multi-fidelity Bayesian optimization literature in terms of the number of independent fidelities as well as decision variables. 
The optimal reactor geometry is 3D printed and experimentally validated using associated optimal operating conditions.

The rest of this article is structured as follows. 
Section \ref{background} provides background around approaches for multi-fidelity Bayesian optimization. 
Section \ref{methodology} outlines our methodology, starting with the problem setting and including details about simulation properties, reactor parameterization and fidelities. 
We then detail our specific approach to the multi-fidelity Bayesian optimization of simulated chemical reactors. 
Section \ref{results} presents our results including experimental validation via 3D printing. 
Finally, Section \ref{conc} outlines our conclusions and future work.

\section{Background}\label{background}

\subsection{Notation}

For the rest of this work, we apply the following notation. 
$\mathbf{x}\in\mathcal{X}\subseteq\mathbb{R}^n$ are decision variables, or inputs, where $\mathcal{X}$ represents the region of feasible inputs.
Similarly, $\mathbf{z}\in\mathcal{Z}\subseteq\mathbb{R}^m$ are fidelity parameters, defined within the continuous or discrete set $\mathcal{Z}$;
$m$ may potentially be greater than one, and we place no restriction on the number of continuous/discrete fidelities available \footnote{Using the traditional naming convention of `multi-fidelity', in the case that $m>1$, approaches could feasibly be referred to as `multi-multi-fidelity' though here we maintain the traditional convention.}.
To make this distinction clearer, when $\mathbf{z}\in \mathbb{R}^1$, it is denoted in `normal' type, $z$.
In the case that $m>1$, $\mathcal{Z}$ is a non-ordered set. 
Therefore, we denote $\mathbf{z}_\bullet$ as the component-wise vector of highest fidelities, following the convention of \citet{mf_continuous}.
We denote the function to be optimized as $f^*$, that takes arguments $\mathbf{x}$ and $\mathbf{z}$ and returns an objective value $y$ with associated computational cost $c$. 
The set of previously evaluated inputs, fidelities, objective values, corresponding to the total number of data $D$ available at a given iteration $t$ is denoted as $\mathcal{D}_t:=\{(\mathbf{x}_i,\mathbf{z}_i,y_i,c_i)\}^D_{i=1}$.
A model of objective $f$ at iteration $t$ is denoted $\hat{f}_t$, and a model of cost at iteration $t$ is denoted $\lambda_t$.
Given that in this work $\lambda$ and $\hat{f}$ will be modeled using Gaussian processes, the mean and standard deviation are denoted as $\mu$ and $\sigma$, respectively, and are indexed by their respective model. 
For example, the mean of the posterior Gaussian process modeling $f$ is denoted $\mu_{\hat{f}_t}$.

\subsection{Gaussian Processes}

A Gaussian process is an infinite-dimension generalization of a multi-variate Gaussian distribution \citep{williams2006gaussian}. 
The mean vector and covariance matrix are replaced by mean and kernel functions, respectively. 
A Gaussian process can be described as
\begin{align*}
    f(x) \sim \mathcal{G}\mathcal{P}(m(\mathbf{x}),k(\mathbf{x},\mathbf{x}^{'})).
\end{align*}
The kernel function $k$ dictates the behavior of functions from this distribution, and can be parameterized by hyper-parameters including length scale.
By conditioning a Gaussian process on a dataset $\mathbf{X}_*$, a posterior distribution of functions can be obtained. 
At inputs $\mathbf{X}$, and previously evaluated function values $\mathbf{y}$ the posterior predictive mean and standard deviation become
\begin{align*}
    \mu_f(\mathbf{X})&=K(\mathbf{X},\mathbf{X}_*)K(\mathbf{X},\mathbf{X})^{-1}\mathbf{y} \\ 
    \sigma_f(\mathbf{X}) &= K(\mathbf{X}_*,\mathbf{X}_*)-K(\mathbf{X}_*,\mathbf{X})K(\mathbf{X},\mathbf{X})^{-1}K(\mathbf{X},\mathbf{X}_*)
\end{align*}
where $K$ is a covariance matrix derived from kernel function $k$.
The ability to derive analytical probabilistic predictions make Gaussian processes an attractive modeling framework.

\subsection{Bayesian Optimization}
Bayesian optimization is a model-based approach for the solution of expensive black-box optimization problems.
The black-box optimization problem is stated as
\begin{align}
    \mathbf{x}^* = \mathop{\mathrm{argmax}}_{\mathbf{x}\in \mathcal{X}}\; f(\mathbf{x}), \label{black_box_problem}
\end{align}
where the black-box function $f$ is evaluated with decision variables $\mathbf{x}$, and bounded by the set $\mathcal{X}$.

A number of approaches can be applied in order to solve Eq.\ref{black_box_problem}.
These include direct methods, which solely rely on function evaluations to inform which value of $\mathbf{x}\in\mathcal{X}$ is selected for evaluation next \citep{larson_menickelly_wild_2019}. 
Alternatively, model-based methods leverage previous function evaluations to learn a model, $\hat{f}$, of $f$.
Given that $\hat{f}$ is often cheaper to evaluate than $f$, and gradient evaluations of $\hat{f}$ are available by design, $f$ can be optimized tractably, and used to inform the selection of the next point. 
Model-based methods are known to be more efficient on a function evaluation basis than direct-methods, and are applied when $f$ is computationally expensive and a large amount of information is required from each evaluation \citep{larson_menickelly_wild_2019,vandeBerg2022,Wang2022,Boukouvala2012}. 

Within Bayesian optimization, a Gaussian process is trained as the surrogate model of $f$ from an initial data set of potential solutions \citep{garnett2023bayesian}. 
The selection of the next point to be evaluated is then made based on a combination of the expected value, and the predicted variance of $f$, as a Gaussian process provides a posterior probability distribution of function values.
Equation \ref{UCB} demonstrates the Upper Confidence Bound (UCB) criterion for selecting the next sampled decision variable $x_{t+1}$:
\begin{align}
    \mathbf{x}_{t+1} = \mathop{\mathrm{argmax}}_{\mathbf{x}\in\mathcal{X}} \; \mu_{\hat{f}_t}(\mathbf{x}) + \beta^{1/2}\sigma_{\hat{f}_t}(\mathbf{x}), \label{UCB}
\end{align}
where $\beta^{1/2}$ is a hyper-parameter controlling the trade-off between exploration ($\sigma$) and exploitation ($\mu$). 
Other acquisition functions have been proposed and applied including probability of improvement \citep{Shahriari2016}, expected improvement \citep{snoek2012practical}, and entropy search \citep{hernandez2014predictive}.
\citet{garnett2023bayesian} provides an excellent overview into alternative Bayesian optimization approaches.

Recently, domain-specific adaptations to Bayesian optimization have been developed, leading to improved performance in specific applications \citep{Na2021,Gonzlez2023,snake}. 
For example, alternative formulations have been proposed to facilitate the setting where function evaluations can be made in parallel resulting in larger information gain and potentially faster convergence \citep{Gonzlez2023,Shahriari2016,Nguyen2019}.
Other domain-specific adaptations to Bayesian optimization include approaches for scenarios where there is an additional cost incurred for evaluating decision variables that are `far' from previous evaluations \citep{snake}, various approaches for high-dimensional \citep{NEURIPS2019_6c990b7a}, and multi-objective \citep{pmlr-v180-daulton22a} Bayesian optimization have also been developed. 

\subsection{Multi-fidelity Bayesian Optimization}

Multi-fidelity approaches take advantage of one or more cheaper to evaluate, but potentially biased and more uncertain, function evaluations. 
Correctly leveraging this information can lead to accelerating design and optimization.
In a multi-fidelity context, the black-box optimization problem presented in Eq.\ref{black_box_problem} becomes 
\begin{align}
    \mathbf{x}^* = \mathop{\mathrm{argmax}}_{\mathbf{x}\in \mathcal{X}}\; f(\mathbf{x},\mathbf{z}_\bullet), \label{mf_bbp}
\end{align}
where fidelity parameters $\mathbf{z}$ may influence both the accuracy and cost, computational or otherwise, of a function evaluation.
Equation \ref{mf_bbp} highlights the emphasis on optimizing for the highest fidelity $\mathbf{z}_\bullet$, given this is the system considered `most true'.
Multi-fidelity Bayesian optimization algorithms to solve Eq.\ref{mf_bbp} have been developed and applied across a number of domains \cite{mf_continuous,He2017,Stroh2021,FLCH,savage2022,Petsagkourakis2021,Winter2023,Batra2019}.
These approaches all have a common characteristic in that both $\mathbf{x}$ and $\mathbf{z}$ need to be selected at each iteration.
How these decisions are made, with respect to the trade off between information gained, and expense incurred, are key aspects of multi-fidelity Bayesian optimization algorithms. 

\citet{savage2022} apply an algorithm in a setting where a single discrete fidelity parameter dictates the cost and information trade-off, applying a deep Gaussian process (DGP) to model the objective at each discrete fidelity. 
The DGP formulation, and other multi-fidelity Bayesian optimization formulations that rely on sequential multi-fidelity models are inherently limited to discrete fidelities. 
Likewise, due to modeling limitations, multiple independent fidelities cannot be considered.
\citet{savage2022} make the recommendation that when using a discrete multi-fidelity Bayesian optimization method, the number of fidelities should be kept to 2 or 3. 
Increasing the number of fidelities available to the algorithm may result in a more complex and difficult to train multi-fidelity model, with the potential for a less accurate description of the design space. 
In addition, simulations have to be performed at every discrete fidelity in order to gain an initial data set for optimization. 
As the number of discrete fidelities increases more initial simulations have to be performed, reducing the efficiency of the overall algorithm.

\citet{mf_continuous} assume that both inputs and fidelities are smoothly varying in a continuous space, modeling both within a joint space $\mathcal{Z} \times \mathcal{X}$, using a single Gaussian process. 
As a result the modeling approach serves as a continuous or continuous approximation of $\mathbb{Z}$, not limiting the setting to discrete fidelities.
By modeling multi-fidelity data within a joint space, fidelities are able to influence the prediction of objective at other fidelities. 
A two-step approach is applied to select $\mathbf{x}_{t+1}$ and $\mathbf{z}_{t+1}$. 
First, decision variables are selected by performing a standard UCB step, conditioned at the highest fidelity.
To select $\mathbf{z}_{t+1}$, the authors create a set of candidate fidelities $\mathcal{Z}_{t+1}$.
From this set, the lowest cost fidelity is selected as $\mathbf{z}_{t+1}$.  
This approach can optimize functions parameterized by an arbitrary number of continuous fidelities. 
However, as the number of fidelities increases the selection of candidate fidelities $\mathcal{Z}_{t+1}$ becomes more difficult as the size of $\mathcal{Z}$ grows exponentially. 
Additionally, the approach assumes that the cost of a simulation at a given fidelity, $\lambda(\mathbf{z})$ is independent of $\mathbf{x}$.

\citet{He2017} assumes that the cost of a function evaluation is both a function of $\mathbf{x}$ and $\mathbf{z}$, presenting the case study of a finite-element thermal and fluid dynamics model of alloy casting with each simulation takes approximately 10 minutes to complete.
To solve the multi-fidelity optimization problem, the authors present an augmented acquisition function, weighted by the cost of a simulation. 
However, all case studies considered are 2, 3, or 4 dimensional, and each case study consists only of a single discrete fidelity parameter. 
Similarly, in obtaining results the authors relax the assumption that $\mathbf{x}$ influences $\lambda$.
\citet{thodoroffmulti} presents a cost-adjusted criterion to select both $\mathbf{x}_t$ and $\mathbf{z}_t$ in a single step, in the context of ice-sheet simulation experimental design. $\mathbf{x}$ and $\mathbf{z}$ are modeled in a joint space by a single Gaussian process $\hat{f}(\mathbf{x},\mathbf{z})$.
Simulation cost is a separately modeled function of both $\mathbf{x}$ and $\mathbf{z}$, with $\mathbf{x}_t$ and $\mathbf{z}_t$ are chosen in a single step. 
Experimental design criteria are modified to weight expected information gain by simulation cost, and both decision variables and fidelity are selected within the same step, necessary if cost $\lambda$ is a function of both $\mathbf{x}$ and $\mathbf{z}$.
\citet{Serani2019} applies multi-fidelity optimization within the CFD domain to benchmark problems including airfoil and boat hull design.
\citet{Huang2006} applied a Bayesian optimization-based technique to a sequentially-trained multi-fidelity model to optimize a metal-forming finite-element simulation.
\citet{march_willcox_wang_2011} perform multi-fidelity optimization of a typical airfoil design problem using a trust-region approach.
\citet{Mansour2020b} optimize the geometry of a coiled tube reactor without pulsed-flow conditions using a genetic algorithm approach. The work assumes a single simulation fidelity, with a result taking over two weeks to produce.

In this article, we build upon previous work for the optimal design and operation of simulated chemical reactors. 
Our contributions are three-fold.
Firstly, we present a novel framework (\textsc{DARTS}) for the multi-fidelity Bayesian optimization of expensive reactor and tube simulations.
Our methodology chooses a simulation fidelity at each iteration based on a cost-adjusted estimated information density, with the overall aim to optimize for the highest fidelity given a time budget.  
In addition, it utilizes a novel stopping criteria, ensuring that a high-fidelity solution is returned.
Secondly, we apply our framework to the industrially relevant problem of simultaneously optimizing the geometry and operating conditions of a simulated pulsed-flow helical-tube reactor.
Unlike the majority of approaches, our framework allows for the cost of a simulation to be modeled as a function of both simulation fidelities and decision variables. 
We also account for two independent fidelities, the axial and radial cell count of a simulation, both of which are treated as jointly continuous. 
Thirdly, we validate assumptions made within the methodology using experimental data. 
The optimal geometry is 3D printed and evaluated using associated optimal pulsed-flow operating conditions, validating both the performance of the framework as well as our multi-fidelity model.
Our approach is extensible to a large number of potential design and optimization problems, ranging from microscale to industrial reactors.
We make all the code available to facilitate reproducibility. 

\section{Method}\label{methodology}

\subsection{Parameterization}

\begin{figure}[htb!]
    \centering
    \includegraphics[width=0.8\textwidth]{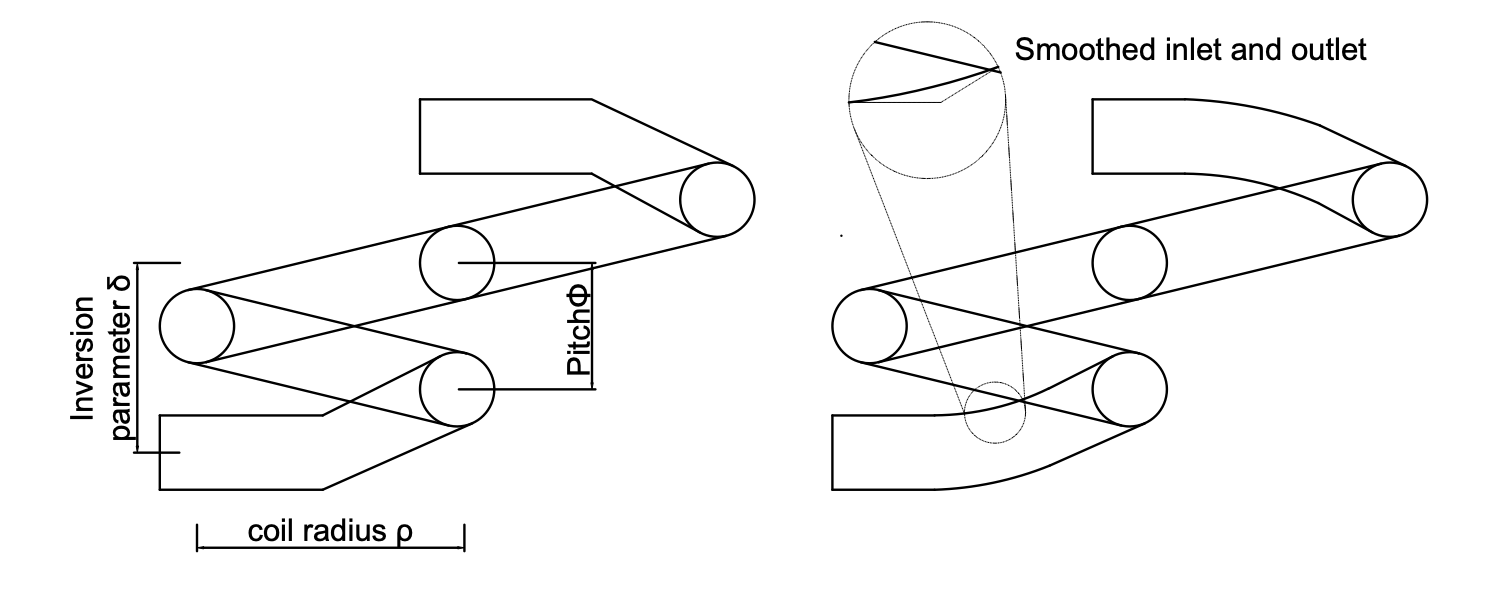}
    \caption{\textit{Left}: A side-view of how coil radius, pitch, and the inversion parameter effects a given coil, with an additional horizontal inlet and outlet. \textit{Right}: When parameterizing a coil with horizontal inlet and outlet, we include a smooth transition based on a quadratic interpolation of points.}
    \label{fig:coil_parameterization}
\end{figure}

A helical-tube reactor is parameterized by a coil radius, coil pitch, and inversion.
Coil pitch denoted by $\phi$ controls how extended the helical tube is, coil radius denoted by $\rho$ controls how tight the coils are within the helical tube, and the inversion parameter is denoted by $\delta$ controls the change in coil direction.
Inversions within helical-tube reactors have been shown to provide effective mixing properties by\citet{McDonough2019a,Rossi2017,Singh2016};
$\delta$ takes a value between 0 and 1, and specifies where along the coil the inversion takes place. 
The length of the coil is maintained as fixed, resulting in all parameterized coils having the same volume.
Within the parameterization, we include a fixed-length inlet and outlet to the coil.
The inlet and outlet are horizontal, and a smooth interpolation is used to ensure that the transition from inlet to coil and coil to outlet, is smooth. 
Figure \ref{fig:coil_parameterization} demonstrates the parameterization we apply within this work. 

The reactor inlet flow is at a Reynolds number of 50 for which relatively insignificant mixing is expected to take place. A superimposition of oscillatory velocity is, therefore, needed to operate under a wide range of plug flow conditions \cite{McDonough2019a}. 
This oscillatory velocity $v_o$ is achieved through parameters representing oscillation amplitude $a$ and frequency $f$.
Oscillatory velocity is defined as 
\begin{align*}
    v_o= 2\pi fa\sin(2\pi ft).
\end{align*}
We vary oscillatory parameters along with the design parameters within the optimization procedure.
 
\subsection{Fidelities}

In addition to geometric design and oscillatory parameters, the output of a simulation is also influenced by one or more fidelities $\mathbf{z}$. 
A typical fidelity used within CFD simulations is the number of discrete finite elements that are contained within the mesh, known as cell count.
Fluid behavior within pulsed-flow helical-tube reactors is complex due to the transient formation and unravelling vortex structures affecting axial and radial mixing. 
Therefore, it is important to capture simulation resolution both axially, through the length of the tube, as well as radially throughout the cross-section of the tube for a high-quality simulation.
In this work, as opposed to applying cell count as a single scalar fidelity parameter, we instead maintain the ability to adapt the resolution of a simulation both axially and radially resulting in two independent fidelities.
We implement a custom meshing procedure that allows for axial and radial fidelity to be varied independently.
The simulation that serves as the highest fidelity $\mathbf{z}_\bullet$ corresponds to when both the axial and radial fidelities are greatest. 
Whilst both axial and radial fidelities are discrete due to their meaning within a finite element context, we treat them as continuous parameters.
During mesh creation values of both fidelities are rounded to the nearest integer.
Within the solution to CFD problems, a number of solver options may be considered as valid fidelities. 
However, as many of these take on non-continuous or Boolean values, we leave their integration for future work. 

\begin{figure}[htb!]
    \centering
    \includegraphics[width=0.8\textwidth]{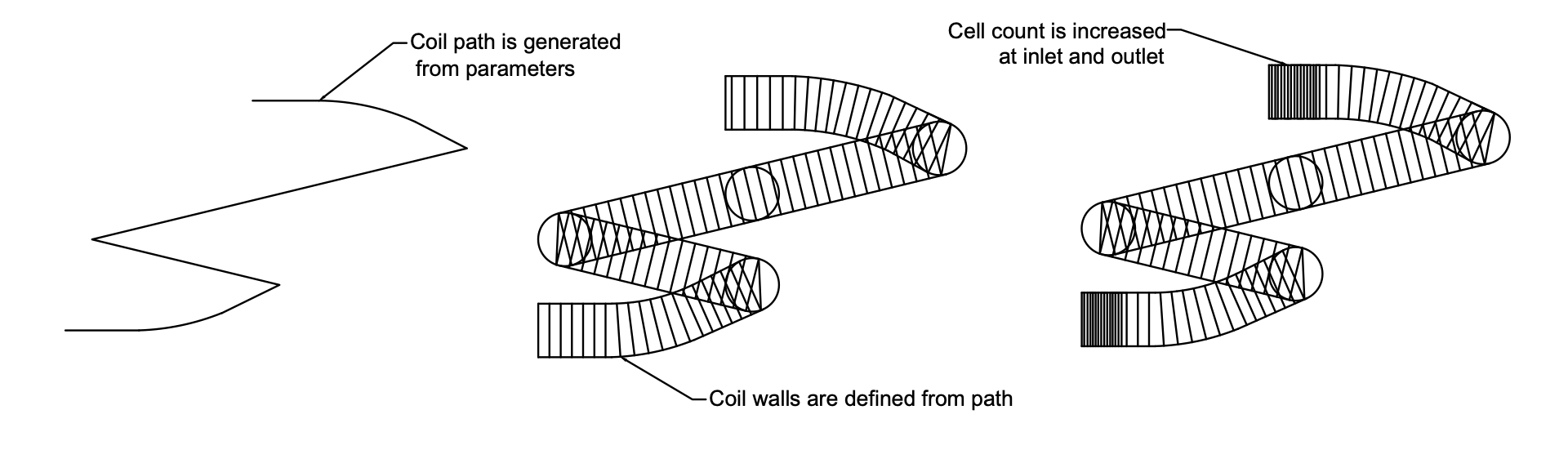}
    \caption{A side-view of the meshing procedure of a coil based on axial fidelity. We first define a central path based on geometric parameters. Subsequently, circles (tube cross-sections) are defined along this path the number of which depends on the axial fidelity. Finally, the mesh density near the inlet and outlet walls is increased by raising the number of circles to track the tracer concentration with higher accuracy.}
    \label{fig:coil_fidelities}
\end{figure}

\begin{figure}[htb!]
    \centering
    \includegraphics[width=0.8\textwidth]{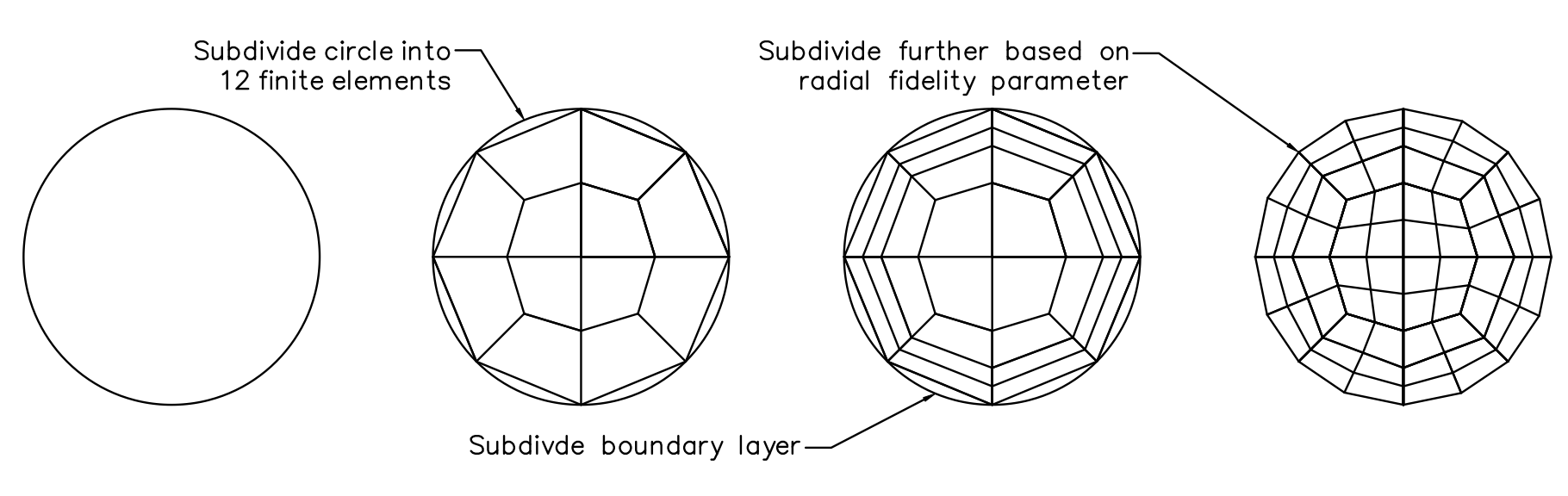}
    \caption{A cross-section of the mesh radially. First, a coarse finite element topology is created. Subsequently, the number of cells near the tube wall are increased in order to better capture the effects of fluid flow near the boundary layer. Finally the mesh is further subdivided based on the radial fidelity value.}
    \label{fig:coil_fidelities_axial}
\end{figure}

Figure \ref{fig:coil_fidelities} demonstrates the final mesh affected by axial fidelity, as well as providing an outline of how meshing is performed. 
Increased mesh density near the inlet and outlet wall is induced throughout all radial fidelity values to reduce numerical diffusion errors that can compromise the accuracy of the CFD solution. 
Figure \ref{fig:coil_fidelities_axial} demonstrates how the final mesh is affected by radial fidelity, as well as providing an outline of the meshing procedure through the creation of a block-structured mesh. 
We refine the mesh  near the walls to better capture the boundary layer and thereby velocity gradients affecting flow characteristics.

\subsection{Simulation}
To perform an evaluation of a given reactor mesh and set of operating conditions, a simulation is performed using the open-source code  OpenFOAM \citep{jasak2007openfoam}. 
An impulse tracer is injected as a scalar field at the inlet of the reactor until $t=0.15$ s into the water as a medium.  
The concentration of the tracer is tracked by solving a convection-diffusion equation through scalarTransportFoam. 
Here, the diffusion coefficient is set as a constant with a small value of \textit{1$\times$$e^{-10} \ m^2/s$} as diffusion in liquids is slow and is dominated by convection. 
The pressure-velocity coupled, transient pimpleFOAM solver is used for solving the unsteady momentum equations as time-dependent oscillatory velocities are introduced.
The pimpleFoam solver is integrated with the scalarTransportFoam through `Solver function Objects'. 
The convection flux on the computational cells was calculated using second-order discretization schemes to ensure the numerical accuracy of the solution. 
The groovyBC boundary condition is used for imposing oscillatory velocity through swak4Foam library \citep{gschaider2013incomplete}. 
This oscillatory velocity along with the steady velocity was initialized at the inlet as Hagen-Poiseuille parabolic velocity profile to cut down on the coil length needed for the flow development and the computational cost.  
Additionally, we terminate the solution by monitoring the tracer concentration at the outlet; this is processed when the tracer concentration drops to values less than 1$\times$$e^{-7}$ for 10 consecutive iterations. 
This variable, early-stopping criterion based on output accelerates the optimization procedure, unlike other studies where a fixed termination based on certain iterations is enforced \citep{Mansour2020b}, but introduces another time dependence on the simulation cost. 
This OpenFOAM solver is integrated with the optimization algorithm via the PyFOAM Python library. 

The output from a simulation returned from PyFOAM is a set of concentration values and respective times at the outlet of the reactor. 
This represents the residence time distribution of the reactor. 
To convert this distribution to a single optimization objective, the distribution is transformed to an equivalent number of tanks-in-series, $N$. 
First, the time and concentration values are converted to dimensionless quantities, $\theta$ and $E(\theta)$, respectively. 
Equation \ref{etheta} shows how $E(\theta)$ can be represented as a function of $\theta$ and $N$ \citep{McDonough2019a}:
\begin{align}
    E(\theta) = \frac{N(N\theta)^{N-1}}{(N-1)!}e^{-N\theta}.\label{etheta}
\end{align}
A fitting procedure is performed to obtain a value of $N$ from $E(\theta)$ and $\theta$. 
This may be done based on least-squares error ($L_2$-norm), however, we find that due to non-idealities and assumptions made to derive Eq. \ref{etheta}, using the difference in maximum predicted value and the maximum value returned from the simulation results in a more robust fit. 
The estimated tanks in series, $N^*$, is therefore calculated as
\begin{align}
        N^* = \arg\max_N \; \left|\max\left[{E(\theta)}\right] - \max\left[\frac{N(N\theta)^{N-1}}{(N-1)!}e^{-N\theta}\right]\right|. \label{n_cost}
\end{align}

\begin{figure}[htb!]
    \centering
    \includegraphics[width=0.6\textwidth]{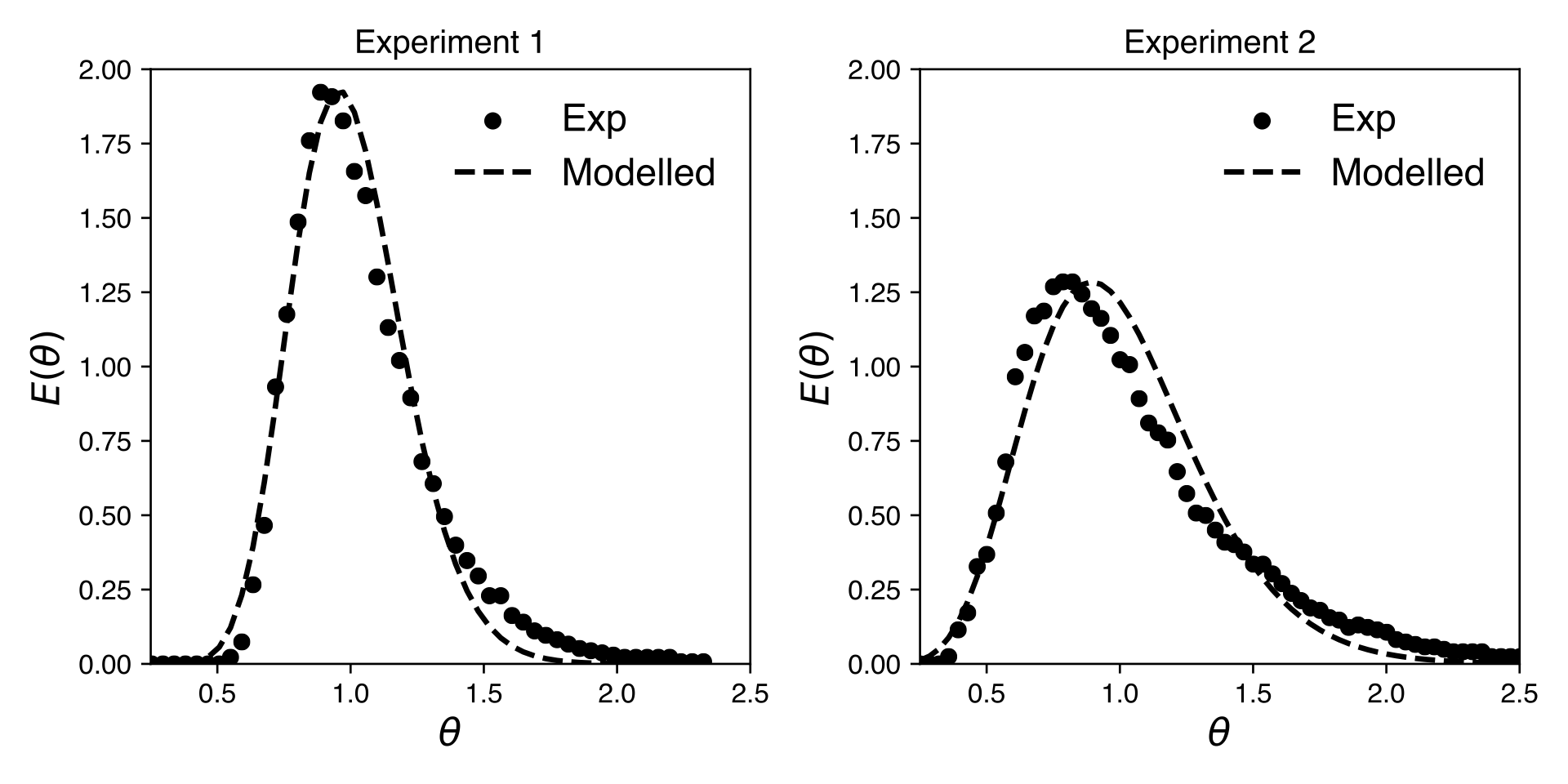}
    \caption{Experimental residence time distribution against the residence time distribution modeled using the ideal tank-in-series relation, given by Eq. \ref{etheta}, with a value of equivalent tanks in series derived from Eq. \ref{n_cost}}
    \label{n_example}
\end{figure}

Figure \ref{n_example} demonstrates $E(\theta)$ against $\theta$ as derived from fitting values of $N$ using Eq. \ref{n_cost}, for two sets of experimental data.

The case file used can be found within the code for this article. 
Though the current case considers the flow with two non-reactive, miscible liquids with identical fluid physical properties, this model can be extended to problems with fluids of differing physical properties modeled as multiphase flows or with reactions where chemical-kinetics can be coupled to CFD simulations in the future.

\subsection{Multi-fidelity Bayesian Optimization}

In addition to dependence on simulation fidelity, the cost of a simulation is also dependent on $\mathbf{x}$ in chemical reactor simulation domains.
Variable step-size CFD solvers, along with early stopping criteria both contribute to how $\mathbf{x}$ can influence simulation cost, particularly in cases where operating conditions are optimized over.
Therefore, the two-step approach presented by \citet{mf_continuous} cannot be applied, where $\mathbf{x}$ is selected first, and then $\mathbf{z}$, as the second-step where information and cost is traded off becomes dependent on $\mathbf{x}$.
Therefore, we follow an approach most similar to \citet{He2017} and \citet{thodoroffmulti}; by utilizing a cost-adjusted acquisition function to trade off information gain and cost in a single step.
To do so, the optimization objectives $\mathbf{y}$ and cost of simulations $\mathbf{c}$ are modeled by two separate Gaussian processes, $\hat{f}$ and $\lambda$, respectively. 
The cost-adjusted acquisition function is as follows:
\begin{align}
    \mathbf{x}_{t+1},\mathbf{z}_{t+1} = \mathop{\mathrm{argmax}}_{(\mathbf{x},\mathbf{z})\in\mathcal{X}\times\mathcal{Z}} \; \frac{ \mu_{\hat{f}_t}(\mathbf{x},\mathbf{z}_\bullet) + \beta^{1/2}\sigma_{\hat{f}_t}(\mathbf{x},\mathbf{z}_\bullet)}{\gamma\mu_{\lambda_t}(\mathbf{x},\mathbf{z})\sqrt{1-k((\mathbf{x},\mathbf{z}),(\mathbf{x},\mathbf{z}_\bullet))^2}}. \label{cost_adjusted}
\end{align}

The numerator on the right-hand side represents expected information gained at the highest fidelity since this is the system of interest ($\mathbf{z} = \mathbf{z}_\bullet$).
The denominator then weights this expected information by the predicted cost of that simulation, enabling lower fidelity solutions to be selected. 
The term $\sqrt{1-k((\mathbf{x},\mathbf{z}),(\mathbf{x},\mathbf{z}_\bullet))^2}$ quantifies how much information is lost when evaluating at a lower fidelity.
The parameter $\gamma$ weights how attractive cheaper solutions are against the information they provide, and $\beta^{1/2}$ weights the exploration-exploitation trade-off as is standard in UCB Bayesian optimization.
However, with a cost-adjusted acquisition function, there is no guarantee that a high-fidelity simulation will be selected for evaluation. 
We address this in the following subsection.

\subsection{Stopping criteria}
In traditional Bayesian optimization, function evaluations are terminated when the computational budget is exhausted, and the evaluated data point with the highest objective is selected as a final solution. 
In a multi-fidelity framework, this criterion contains subtle nuances. 
As the underlying function of interest is $f(\mathbf{x},\mathbf{z}_\bullet)$, the optimal solution returned should be the highest evaluated function-value where $\mathbf{z}=\mathbf{z}_\bullet$.
Throughout optimization $f(\mathbf{x},\mathbf{z}_\bullet)$ is maximized through lower fidelity simulations, with no guarantee that simulations at the highest fidelity will be selected to be evaluated.
To mitigate this detail in their two-step approach \citet{mf_continuous} includes criteria to ensure that in some scenarios $\mathbf{z}_\bullet$ is selected in favor of a lower fidelity.
It follows that when $f(\mathbf{x},\mathbf{z}_\bullet)$ has been optimized `as much as possible' and the most information has been gained, {\it but before all computational budget has been exhausted}, a function evaluation of the highest fidelity needs to take place, to ensure a valid final solution is returned. 
Therefore, we derive the notion of a multi-fidelity stopping criteria for Bayesian optimization. 
The three key criteria for the return of a final highest-fidelity solution are as follows:

\begin{enumerate}
    \item This solution need-not be generated by trading off mean and variance (as within a standard acquisition function), and therefore can be performed greedily given that it will be the final evaluation. 
    \item This solution should be the last solution selected, taking advantage of the maximum amount of information.
    \item There must be enough time remaining to evaluate this solution.
\end{enumerate}

Depending on the inclination of the decision-maker, the third criterion may be relaxed and a final solution may be returned without having been evaluated. 
However, we assume that a valid solution must have been evaluated on the true, high-fidelity function for which $\mathbf{z}=\mathbf{z}_{\bullet}$. 
At each iteration, this greedy, potentially final, solution is obtained by solving the following equation given by

\begin{align}
    \mathbf{x}_g = \arg\max_{\mathbf{x}\in\mathcal{X}} \; \mu_{\hat{f}_t}(\mathbf{x},\mathbf{z}_\bullet).\label{greedyx}
\end{align}

where $\mathbf{x}_g$ is a solution that fulfills the first criterion for a final solution.
In order to ensure that the second and third criteria are adhered to, the cost of evaluating $\mathbf{x}_g$, and the cost of evaluating $\mathbf{x}_{t+1}$ are compared with the time or computational budget remaining. 
As the cost of evaluating a simulation is modeled using a Gaussian process, we apply both the mean and the standard deviation resulting in a probabilistic upper-bound of each evaluation cost.
Equation \ref{t_left} details the maximum time required to evaluate both the next explorative solution as well as the greedy high-fidelity solution
\begin{align}
c_{\max} = \underbrace{\mu_{\lambda_t}(\mathbf{x}_{t+1},\mathbf{z}_{t+1}) + p_\lambda \sigma_{\lambda_t}(\mathbf{x}_{t+1},\mathbf{z}_{t+1})}_{\substack{\text{Predicted maximum cost of}\\\text{next `standard' evaluation}}} +  \overbrace{\mu_{\lambda_t}(\mathbf{x}_g,\mathbf{z}_\bullet) + p_\lambda \sigma_{\lambda_t}(\mathbf{x}_g,\mathbf{z}_\bullet))}^{\substack{\text{Predicted maximum cost of}\\\text{high-fidelity `greedy' evaluation}}}\label{t_left}
\end{align}
where $p_\lambda$ is a parameter weighting the standard deviation, providing the ability to be more conservative and ensure greater probability $\mathbf{x}_g$ can be evaluated.
If the value of $c_{\max}$ is greater than the remaining evaluation budget, then $(\mathbf{x}_g,\mathbf{z}_\bullet)$ should be evaluated as a final solution. 
Otherwise $(\mathbf{x}_{t+1},\mathbf{z}_{t+1})$ should be evaluated as usual, as it is highly probable that there is enough time remaining for both a standard evaluation as well as a greedy evaluation.

In addition, the objective value returned from the solution of Eq. \ref{greedyx}, $\mu_{\hat{f}}(\mathbf{x}_g,\mathbf{z}_\bullet)$, provides a proxy for monitoring the progress of multi-fidelity Bayesian optimization. 
As function values across different fidelities cannot be compared like-for-like, it enables practitioners to observe the progress of optimization. 
Algorithm \ref{STOP} demonstrates our approach for multi-fidelity Bayesian optimization.

\begin{algorithm}
    \KwIn{Remaining budget $\Lambda$, Input bounds $\mathcal{X}$, Fidelity bounds $\mathcal{Z}$, objective function $f$, number of initial samples $D$, exploration parameter $\beta^{1/2}$, cost parameter $\gamma$, kernel function $k$.}
    \KwOut{$\mathbf{x}^*,y^*$}
    
    $\mathcal{D} = \{(\mathbf{x}_i,\mathbf{z}_i,y_i,c_i)\}^D_{i=1}$ \tcp*[f]{create initial data set through DoE}\\
    \While {$\Lambda > 0$}{
     $\hat{f}_t\sim\{\mathcal{GP}\left(\mathbf{0},k((\mathbf{x},\mathbf{z}),(\mathbf{x}',\mathbf{z}')\right)|\mathcal{D}\}$ \tcp*[f]{train objective GP using $\mathcal{D}$}\\
     $\lambda_t\sim\{\mathcal{GP}\left(\mathbf{0},k((\mathbf{x},\mathbf{z}),(\mathbf{x}',\mathbf{z}')\right)|\mathcal{D}\}$ \tcp*[f]{train cost GP using $\mathcal{D}$}\\
     $\mathbf{x}_{t+1},\mathbf{z}_{t+1} \leftarrow$ Eq. \ref{cost_adjusted} \tcp*[f]{solve cost-adjusted UCB}\\
     $\mathbf{x}_g \leftarrow$ Eq. \ref{greedyx} \tcp*[f]{solve greedy high-fidelity problem}\\
     $c_{max} \leftarrow $ Eq. \ref{t_left} \tcp*[f]{estimate maximum budget for both evaluations}

    \eIf {$\Lambda > c_{max}$}{
        $y_{t+1},c_{t+1} = f(\mathbf{x}_t,\mathbf{z}_t)$\tcp*[f]{evaluate cost-adjusted UCB solution}
    }
    {
        $y_g,c_g = f(\mathbf{x}_g,\mathbf{z}_\bullet)$\tcp*[f]{evaluate greedy high-fidelity solution}\\
        $\mathbf{x}^* \leftarrow \mathbf{x}_g$\\
        $y^* \leftarrow y_g$\\
        \Return $\mathbf{x}^*,y^*$
    }
    $\Lambda \leftarrow \Lambda - c_{t+1}$\tcp*[f]{update remaining budget}\\
    $\mathcal{D} \leftarrow \mathcal{D} \cup(\mathbf{x}_{t+1},\mathbf{z}_{t+1},y_{t+1},c_{t+1})$\tcp*[f]{update data set}\\
    }
    \caption{Multi-fidelity Bayesian Optimization for Simulated Chemical Reactors}\label{STOP}
\end{algorithm}

\section{Results}\label{results}

In order to provide a valid comparison between alternative reactor geometries, the length of a given helical coil tube is fixed as 75mm and the tube radius is fixed as 2.5mm. 
This ensures that all reactors within the design space have the same volume, and therefore equivalent number of tanks-in-series can be compared between solutions. 

\subsection{Parameterization}
\begin{figure}[htb!]
    \centering
    \includegraphics[width=\textwidth]{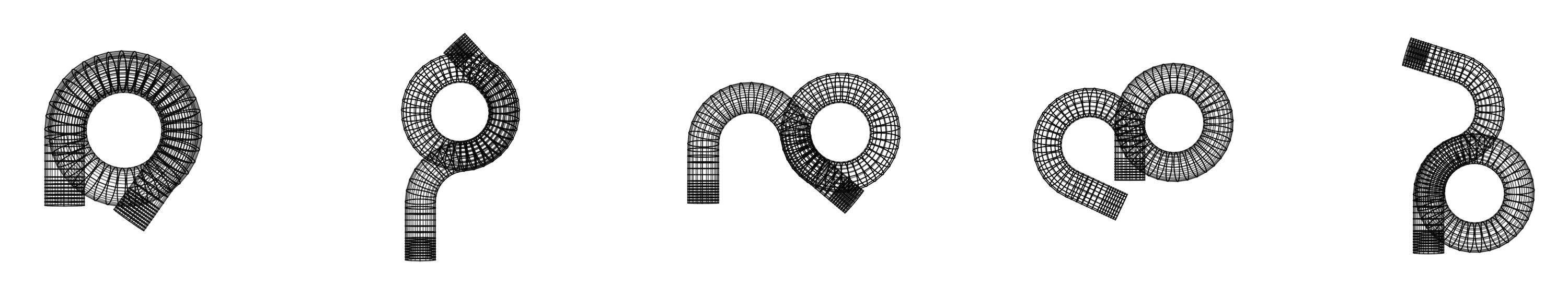}
    \caption{The effect of inversion parameter $\delta$ for a helical coil tube with fixed length, top view. From left to right: $\delta$ is evaluated at 0, 0.15 0.3, 0.6, and 0.75 with coil radius and pitch remaining constant.}
    \label{fig:inversion_demo}
\end{figure}

\begin{figure}[htb!]
    \centering
    \includegraphics[width=\textwidth]{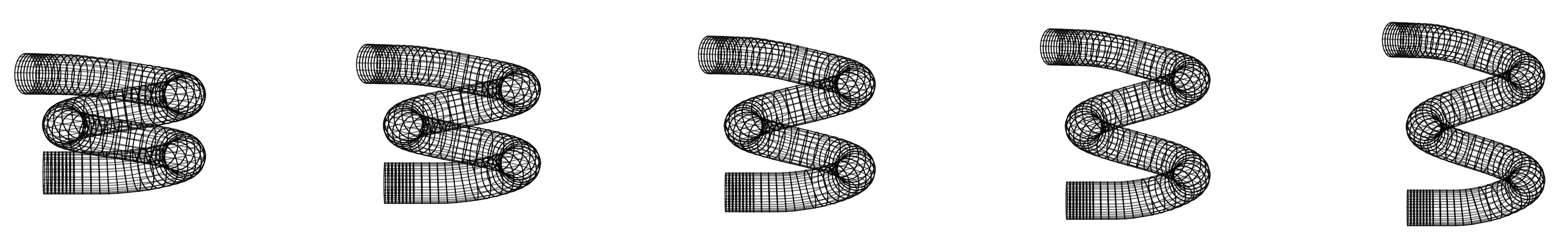}
    \caption{The effect of pitch $\phi$ for a helical coil tube with a fixed length, side view. From left to right: $\phi$ is evaluated from 7.5mm to 15mm with coil radius and inversion location remaining constant.}
    \label{fig:pitch_demo}
\end{figure}

\begin{figure}[htb!]
    \centering
    \includegraphics[width=\textwidth]{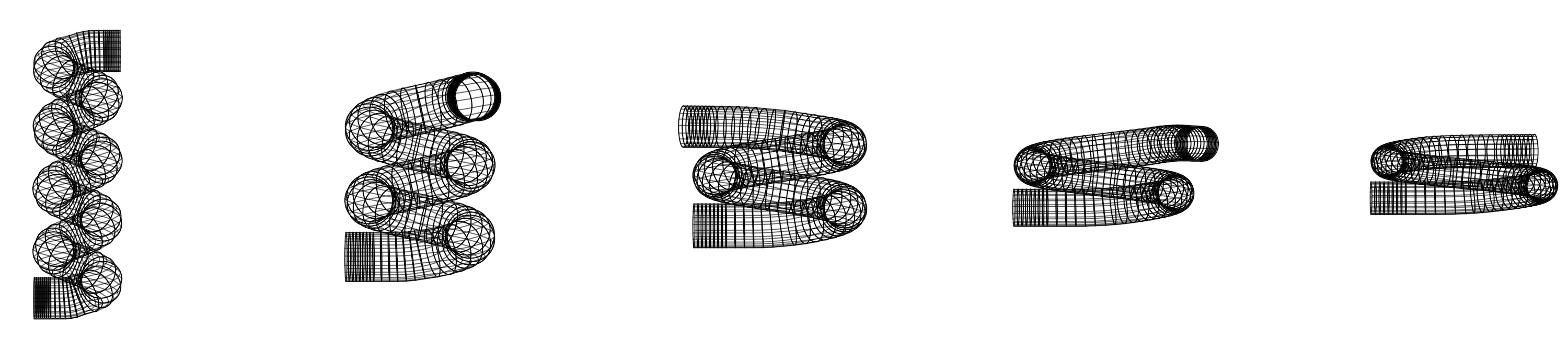}
    \caption{The effect of coil radius $\rho$ for a helical coil tube with a fixed length, side view. From left to right: $\rho$ is evaluated from 3mm to 12.5mm with inversion location and pitch remaining constant.}
    \label{fig:radius_demo}
\end{figure}

Parameters are assigned reasonable bounds based on physically realizable values as well as bounds dictated by the constraints of the experimental setup.
The parameters $\delta$, $\phi$, and $\rho$ are bounded between 0-1, 7.5-15mm, and 3-12.5mm, respectively, while the pulsed-flow amplitude, frequency, and the Reynolds number are bounded between 1-8mm, 2-8 Hz, and 10-50, respectively. 
These bounds ensure that tube self-intersections cannot occur which would have resulted in an unconstrained problem.  
Figures \ref{fig:inversion_demo}, \ref{fig:pitch_demo}, \ref{fig:radius_demo} demonstrate the effect of $\delta$, $\phi$, and $\rho$ on the helical coil tube geometry, respectively.
The mesh for a given set of parameters and fidelities is generated using a custom mesh generation scheme in Python using the \texttt{classy\_blocks} library, available at \url{https://github.com/OptiMaL-PSE-Lab/pulsed-reactor-optimization/}.

\begin{figure}[htb!]
    \centering
    \begin{subfigure}[b]{0.465\textwidth}
         \centering
         \includegraphics[width=0.32\textwidth]{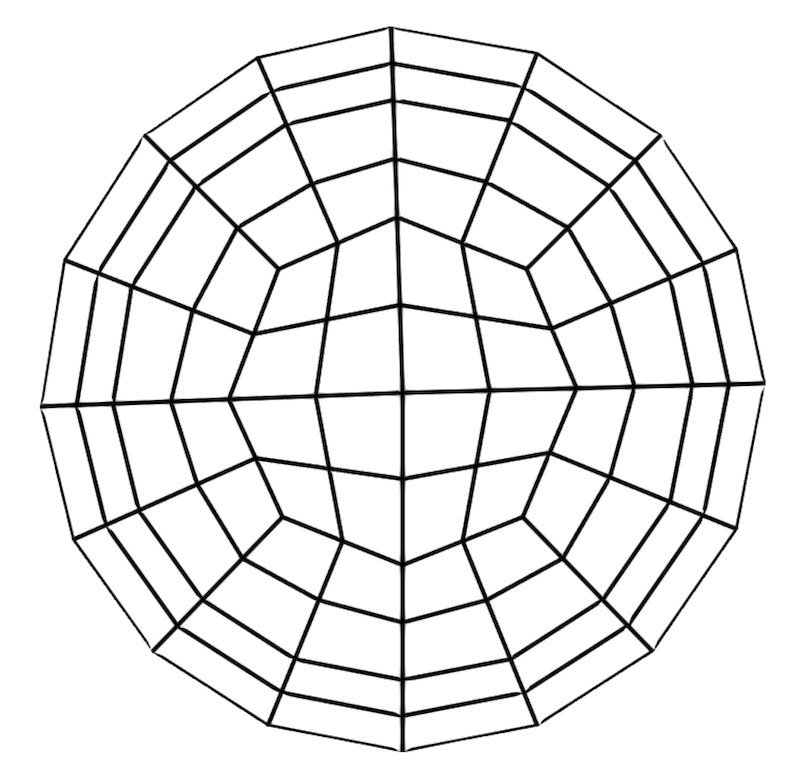}
         \includegraphics[width=0.32\textwidth]{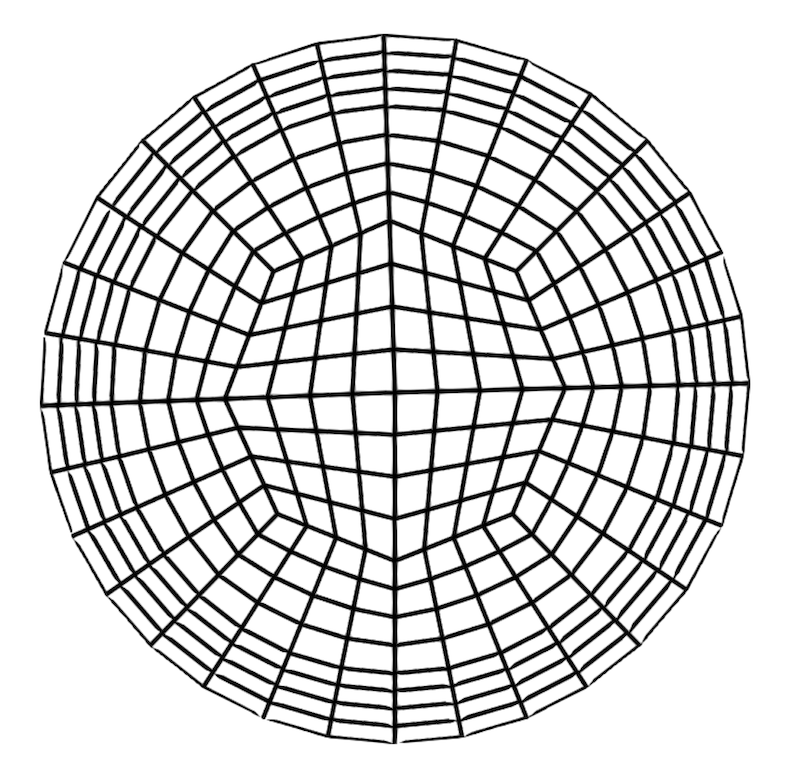}
         \includegraphics[width=0.32\textwidth]{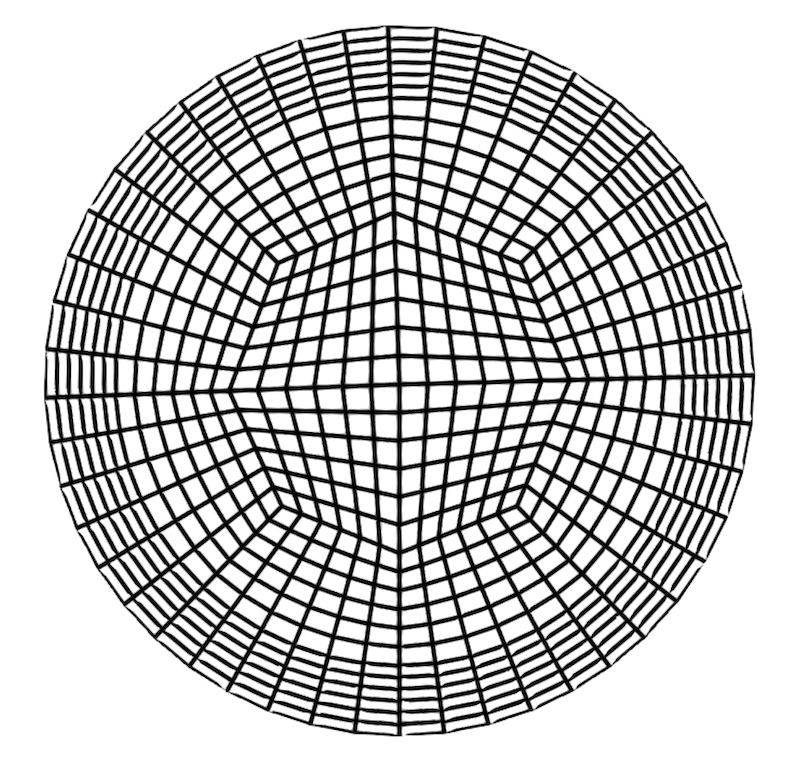}
         \caption{\textit{Radial cross section depicting radial fidelity at values of 1, 3 and 5.}}
         \label{fig:left_radial}
     \end{subfigure}\hspace{2em} 
    \begin{subfigure}[b]{0.465\textwidth}
         \centering
         \includegraphics[width=0.32\textwidth]{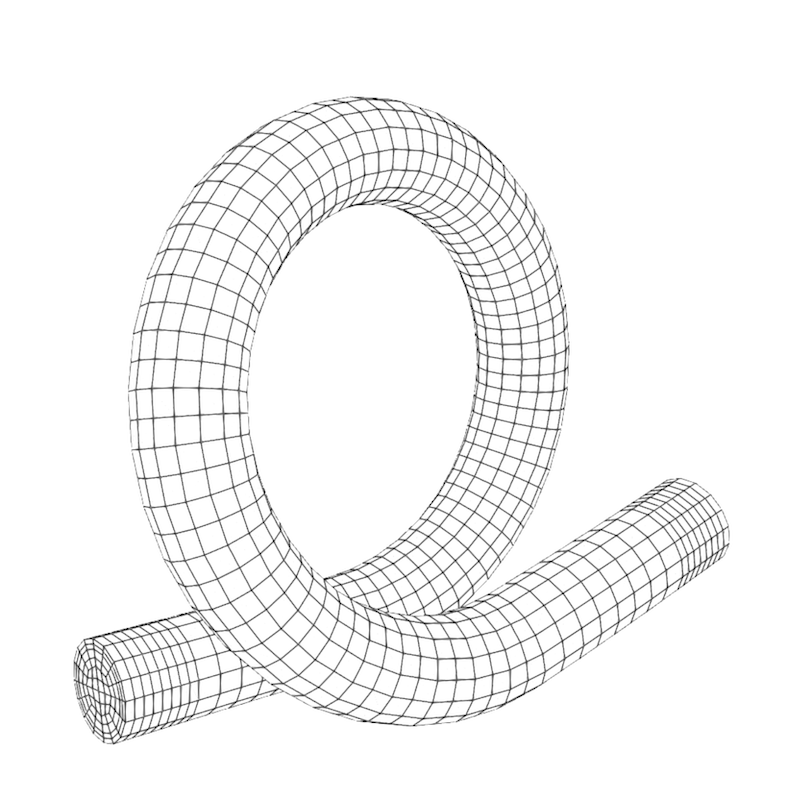}
         \includegraphics[width=0.32\textwidth]{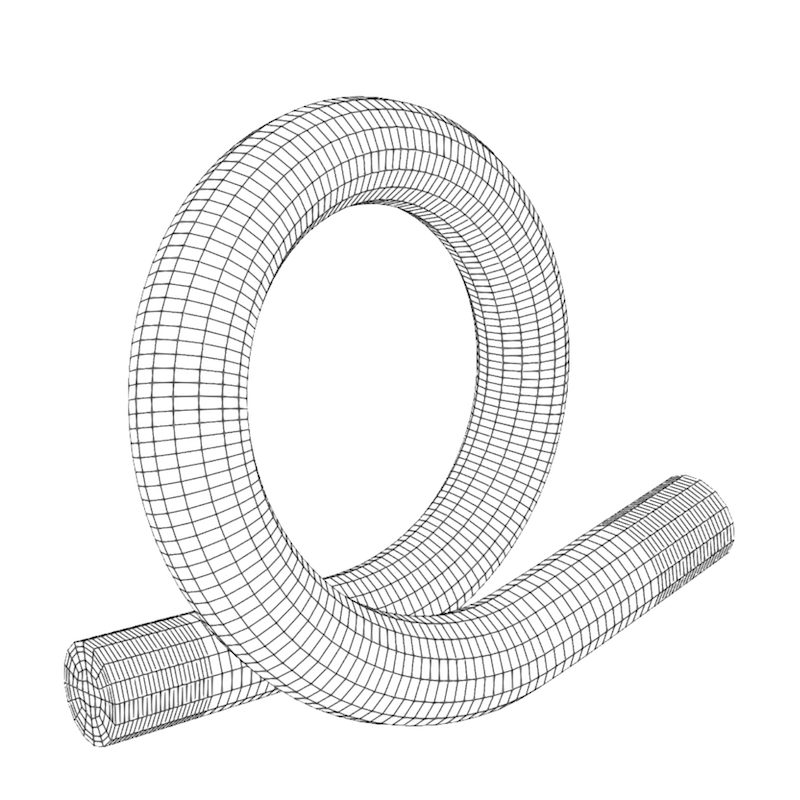}
         \includegraphics[width=0.32\textwidth]{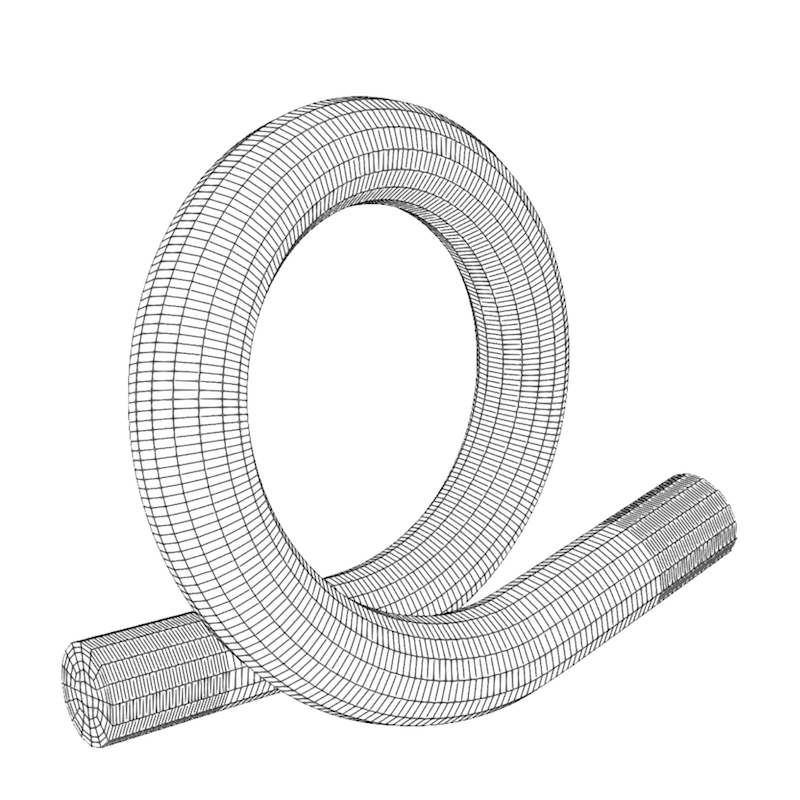}
         \caption{\textit{View of complete reactor mesh at axial fidelity values of 20, 40, and 60.}}
         \label{fig:right_axial}
     \end{subfigure}
     \caption{An instance of helical-tube reactor geometry as affected by axial and radial fidelity.}
     \label{fidelities}
\end{figure}

The lowest axial and radial fidelity values are 1 and 20, and the corresponding highest fidelity values are 5 and 60, respectively.
Figure \ref{fidelities} demonstrates the effect of both axial and radial fidelities on the final mesh, given a fixed set of design parameters.

\subsection{Model Validation}

\begin{figure}[htb!]
    \centering
    \begin{subfigure}[b]{0.465\textwidth}
         \centering
         \includegraphics[width=\textwidth]{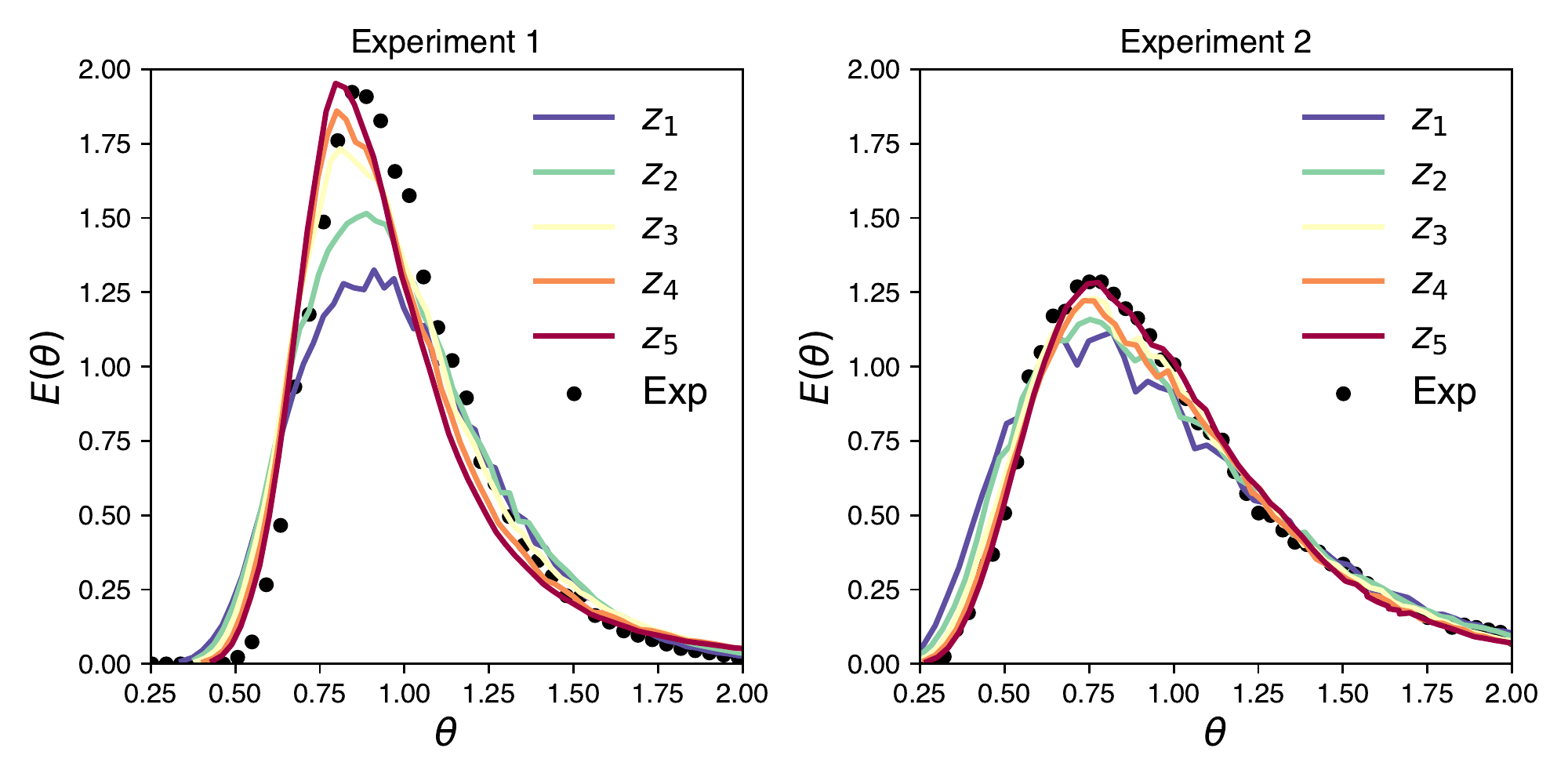}
        \caption{The outlet concentration profile of a tracer impulse at five combinations of fidelity ranging from lowest ($\textbf{z}_1$) to highest ($\textbf{z}_5$) against experimental data. $E(\theta)$ is dimensionless concentration as a function of dimensionless time $\theta$.}   
         \label{fig:left_profile}
     \end{subfigure}\hspace{2em} 
    \begin{subfigure}[b]{0.465\textwidth}
         \centering
         \includegraphics[width=\textwidth]{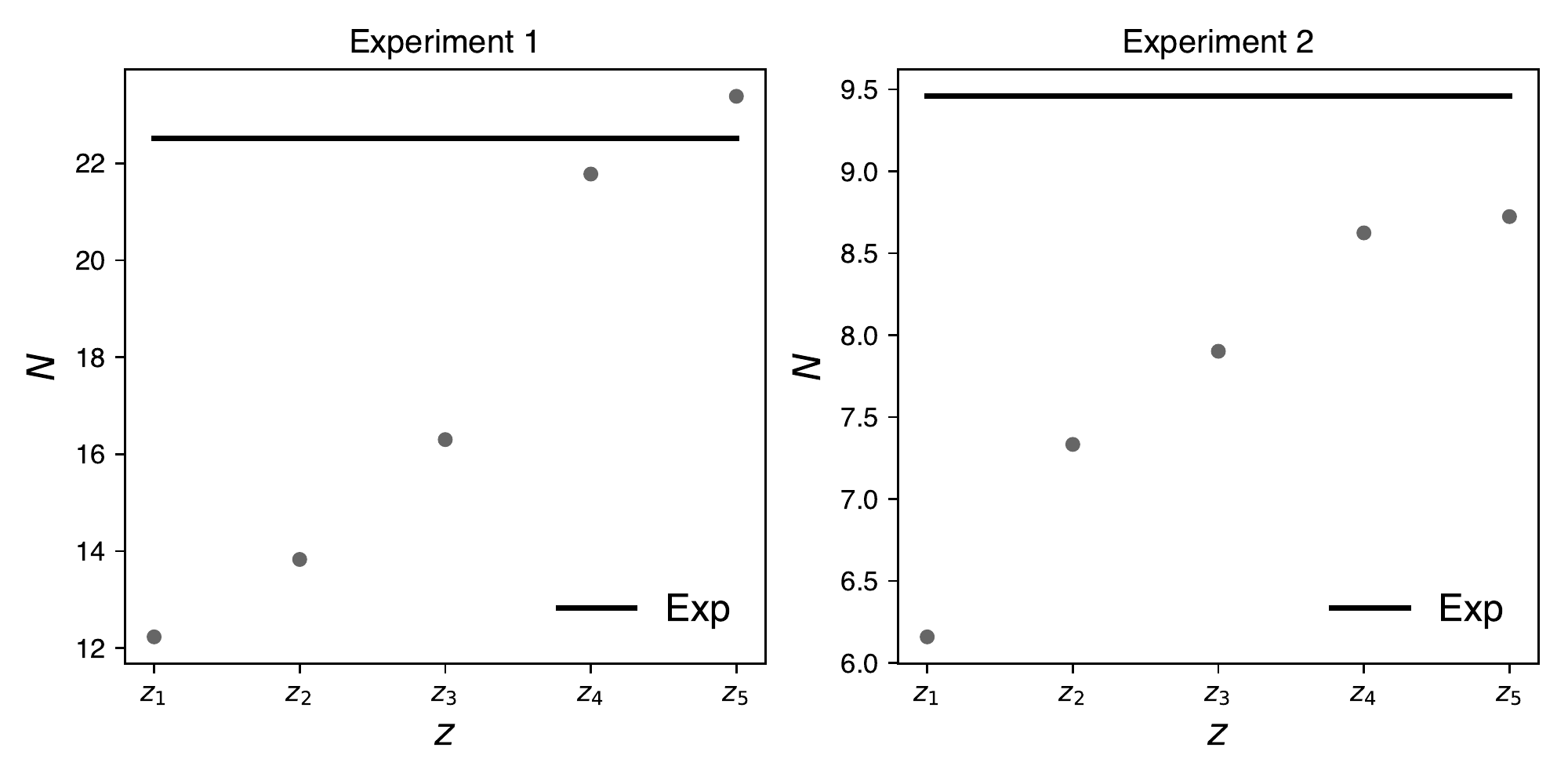}
         \caption{The value of $N$ corresponding to the concentration profile from each fidelity, for each of the five selected fidelity combinations. As overall fidelity is increased from $\textbf{z}_1$ to $\textbf{z}_5$, the experimentally derived value for $N$, $\hat{N}$ is recovered.}
         \label{fig:right_n}
     \end{subfigure}
     \caption{Validation of five discrete mesh fidelities corresponding to different cell counts, across two sets of experimental data under different conditions. }
     \label{validation}
\end{figure}

In order to gain insight into the effect of a simulation fidelity on the accuracy of $f(\mathbf{x},\mathbf{z})$ for a fixed geometry and operating conditions, simulations with five discrete fidelities were performed for fixed $\mathbf{x}$ and compared with experimental data.
In a multi-fidelity context, experimental measurements may themselves be considered the highest-fidelity evaluations available. 
However, we leave this extension for future work and maintain our assumption that the highest fidelity simulation is the function of interest. 
The set of fidelities validated are $\{(z_{\text{axial}}=20,z_{\text{radial}}=1),(z_{\text{axial}}=30,z_{\text{radial}}=2),(z_{\text{axial}}=40,z_{\text{radial}}=3),(z_{\text{axial}}=50,z_{\text{radial}}=4),(z_{\text{axial}}=60,z_{\text{radial}}=5)\}$.
Figure \ref{validation} demonstrates the tracer concentration profile of simulations using these five selected fidelity combinations against two sets of experimental data. 
The experimental data were generated from a 3D-printed reactor with a length of 75.3mm, coil radius of 12mm, coil pitch of 10mm, and no inversion. 
Pulsed-flow operating conditions were induced with a frequency of 5 Hz, the Reynolds number was equal to 50, and the two experiments differed by applying a flow amplitude of 2mm and 4mm. 
An experimental value of equivalent tanks-in-series, denoted by $\hat{N}$, was calculated for both experiments from each resulting residence-time distribution.

Figure \ref{validation} also demonstrates how increasing overall fidelity (and therefore cell count) results in a closer approximation to the experimental value of $N$, derived from each concentration profile.
For both experiments, the nature with which $N$ approaches $\hat{N}$ when the overall fidelity is increased implies that $f(\cdot,\mathbf{z})$ is smoothly varying.
The information that fidelity smoothly influences $f$ validates the assumption that $\textbf{z}\in\mathcal{Z}$ varies smoothly and continuously enabling a Gaussian process to be applied as a multi-fidelity model. 

\begin{figure}[htb!]
    \centering
    \includegraphics[width=\textwidth]{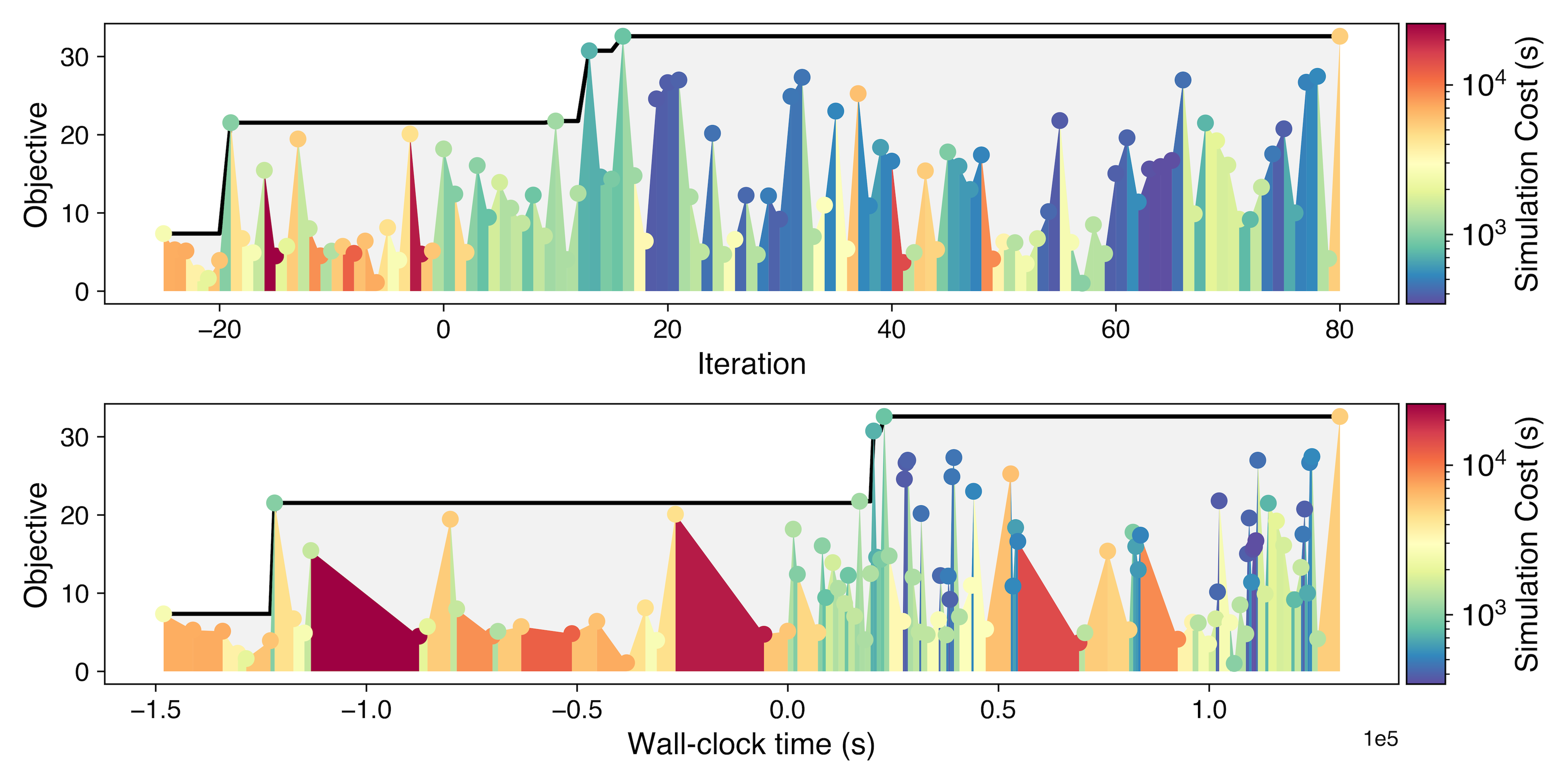}
    \caption{The number of equivalent tanks-in-series evaluated colored by the respective cost of simulation. The upper half of the figure shows these quantities against iteration and the lower half shows these quantities against wall-clock time, highlighting the importance of lower-cost simulations.}
    \label{fig:gamma15_beta25}
\end{figure}

According to the first steps of Algorithm \ref{STOP}, we generate an initial data set of solutions $\mathcal{D} = \{(\mathbf{x}_i,\mathbf{z}_i,y_i,c_i)\}^D_{i=1}$.
To do so, we perform a Latin hypercube design of experiments scheme \citep{McKay1979} to maximize the initial information gained, with 25 solutions ($D=25$). 
This initial number represents a hyper-parameter. 
How best to distribute initial samples across variables and fidelities within multi-fidelity Bayesian optimization, and what constitutes a reasonable number of samples, is yet to be investigated, and we leave this for future investigation. 
We normalise all inputs and outputs to have zero mean and a standard deviation of unity at the beginning of each iteration.
Before solutions are evaluated, we re-scale inputs back to their physical quantities.
Initially, we set the hyper-parameters of Eq. \ref{cost_adjusted} as $\gamma=1.5$ and $\beta=2.5$, and in Eq. \ref{t_left} we set $p_\lambda=2$ resulting in a maximum cost prediction with 95\% confidence. 
The maximum computational budget we specify is 64 hours, not including the budget spent sampling initial solutions, though this may be included. 
Figure \ref{fig:gamma15_beta25} shows the optimization progress throughout the 64 hours.
The time to generate, and objective values of initial solutions generated are denoted by using negative iteration and wall-clock time values. 
Therefore, optimization itself is denoted as starting at iteration 0, and at a wall-clock time of 0. 

Simulations performed to generate the initial data set of solutions are plotted, and are designated to end at iteration and time 0. 
When optimization begins, the search space can be seen to be explored through a number of lower-cost simulations. 
As the value of the objective function at different fidelities is assumed to be biased, the objective values cannot be compared like-for-like. 
However, Fig. \ref{fig:gamma15_beta25} provides insight into the progression of the algorithm.

\begin{figure}[htb!]
    \centering
    \includegraphics[width=\textwidth]{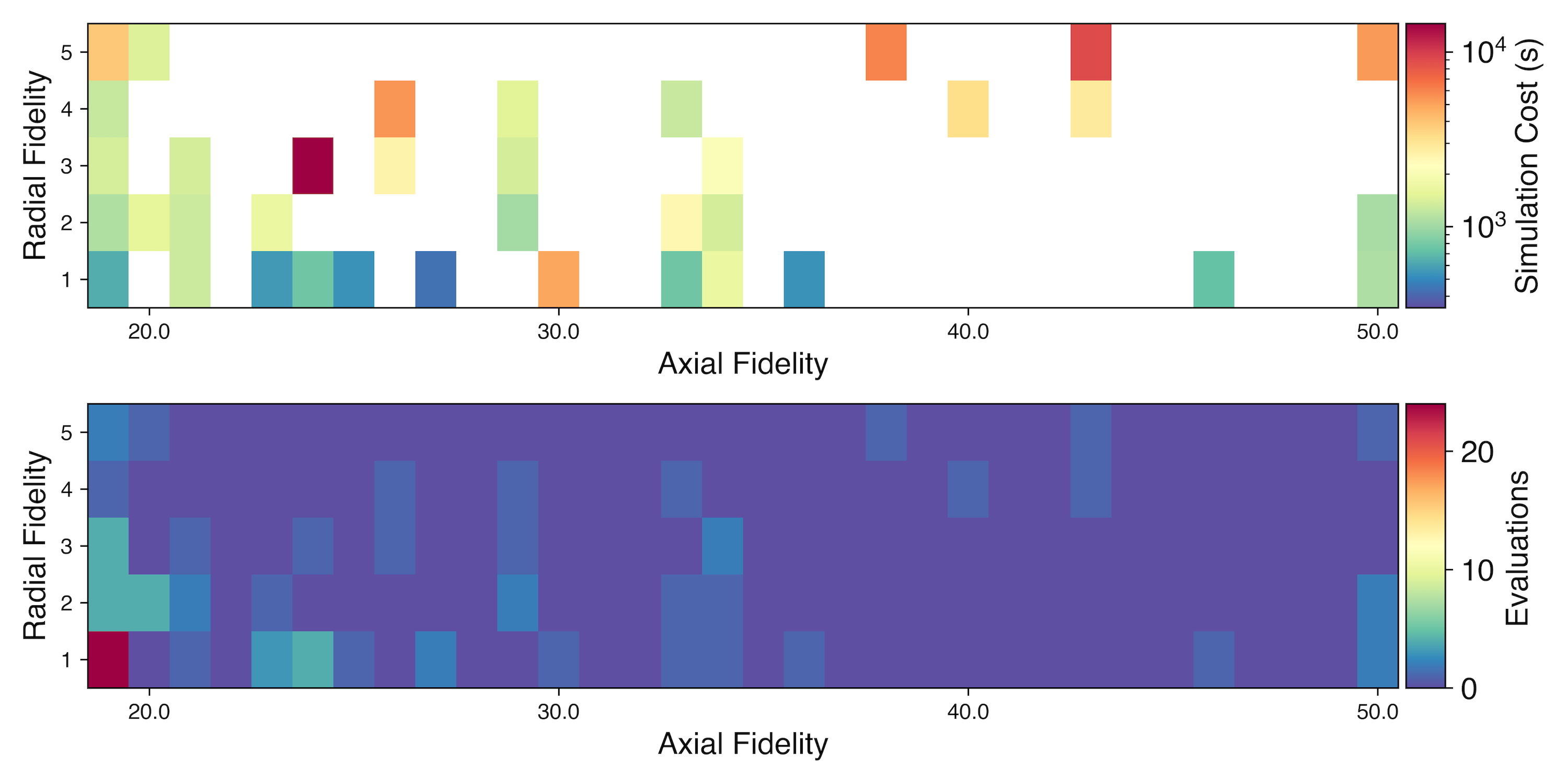}
    \caption{The fidelities selected throughout optimization within $\mathcal{Z}$. The upper plot demonstrates the average simulation cost at each fidelity. As we apply discrete fidelities which are continuously approximated, we present the lower plot which demonstrates the number of times each discrete fidelity is evaluated. Simulation cost is confirmed to be not only a function of fidelity, but $\mathbf{x}$ as well, as there is no clear distribution of simulation costs across $\mathcal{Z}$. }
    \label{fig:fidel}
\end{figure}

\begin{figure}[htb!]
    \centering
    \includegraphics[width=\textwidth]{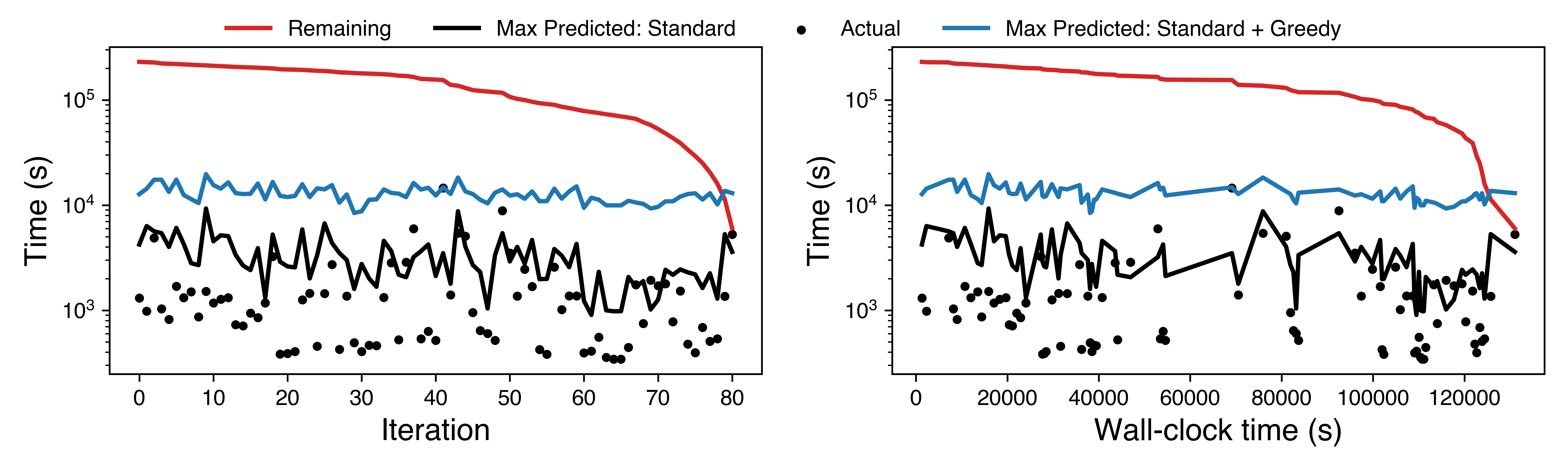}
    \caption{The maximum predicted time to complete a `standard' iteration compared with the maximum predicted time to complete both a standard iteration as well as a greedy iteration, the actual simulation time, and the remaining time. As the blue line passes above the black line, a greedy iteration must be performed as there is not enough remaining budget to perform both a greedy and standard evaluation}
    \label{fig:time_remaining}
\end{figure}

Figure \ref{fig:fidel} shows the combinations of axial and radial fidelity evaluated throughout optimization, alongside respective simulation costs. Generally, simulations performed with lower radial and axial fidelities have lower costs.
However, as this relationship is also dependent on reactor geometry and operating conditions, some simulations have higher costs at lower fidelities. 
The framework tends to evaluate simulations at the lower half of axial fidelity at all five discrete radial fidelity values. 
This indicates that simulations with different axial fidelities are more correlated than simulations with differing radial fidelities.
Less variance across the axial fidelity will result in the acquisition function favouring lower axial fidelity values, as information about the highest fidelity can be gained due to the higher correlation. 
The final high-fidelity solution can be seen at the component-wise highest-fidelity at the top-right of Fig \ref{fig:fidel}.
The quantities tracked throughout optimization in order to ensure this solution is returned are demonstrated in Fig \ref{fig:time_remaining}.

\begin{figure}[htb!]
    \centering
    \begin{subfigure}[b]{0.4\textwidth}
         \centering
         \includegraphics[width=\textwidth]{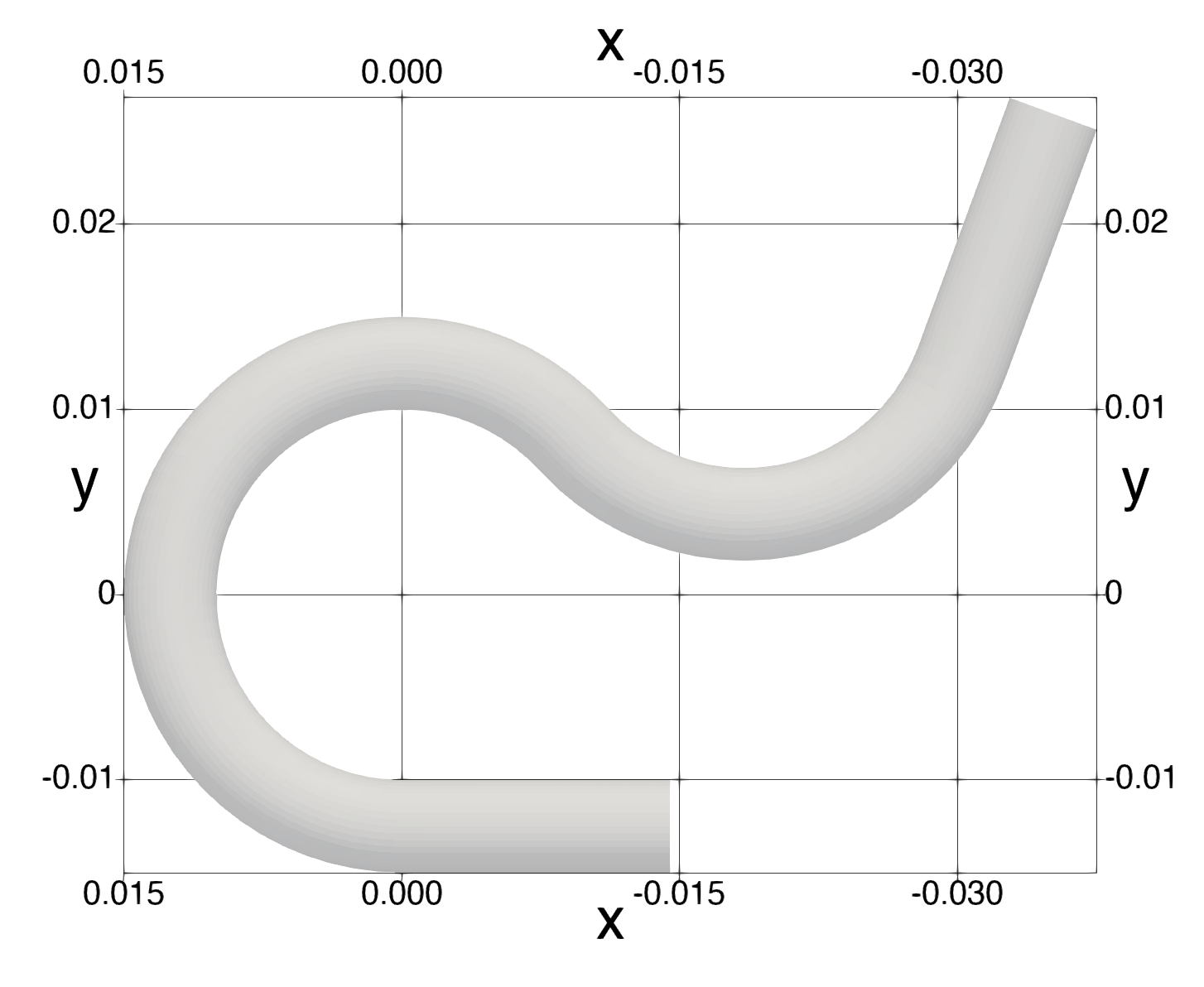}
     \end{subfigure}\hspace{2em} 
    \begin{subfigure}[b]{0.4\textwidth}
         \centering
         \includegraphics[width=\textwidth]{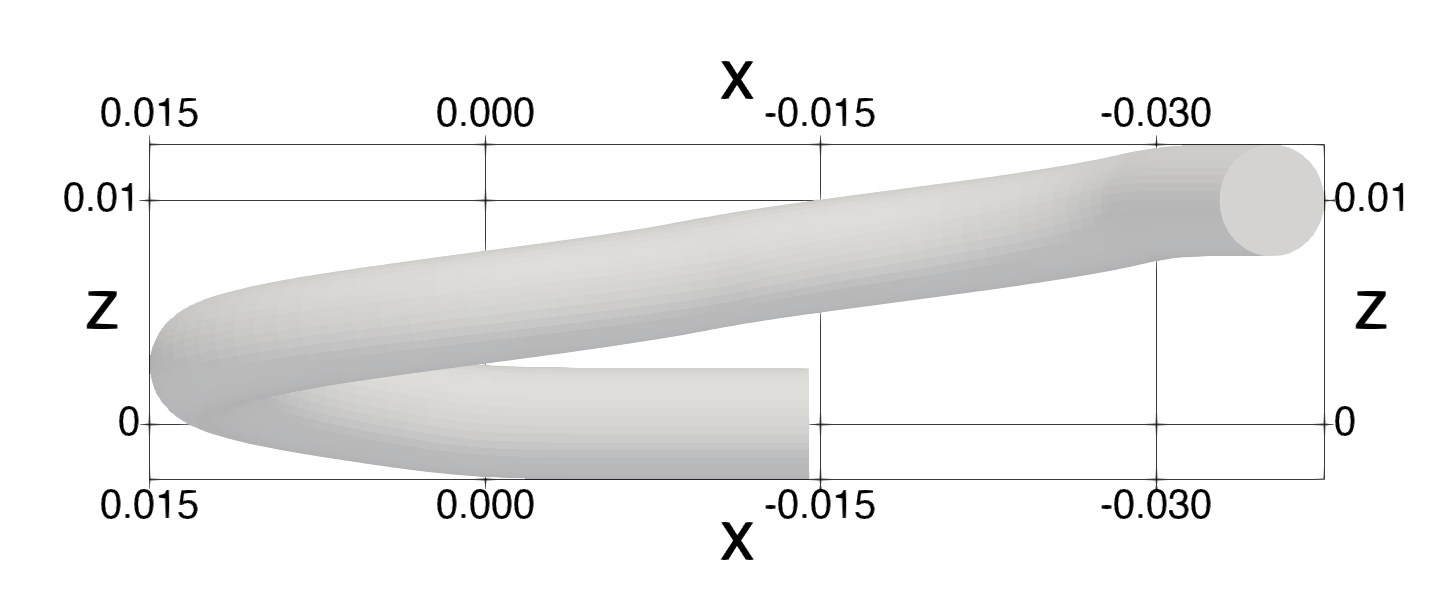}
         \includegraphics[width=\textwidth]{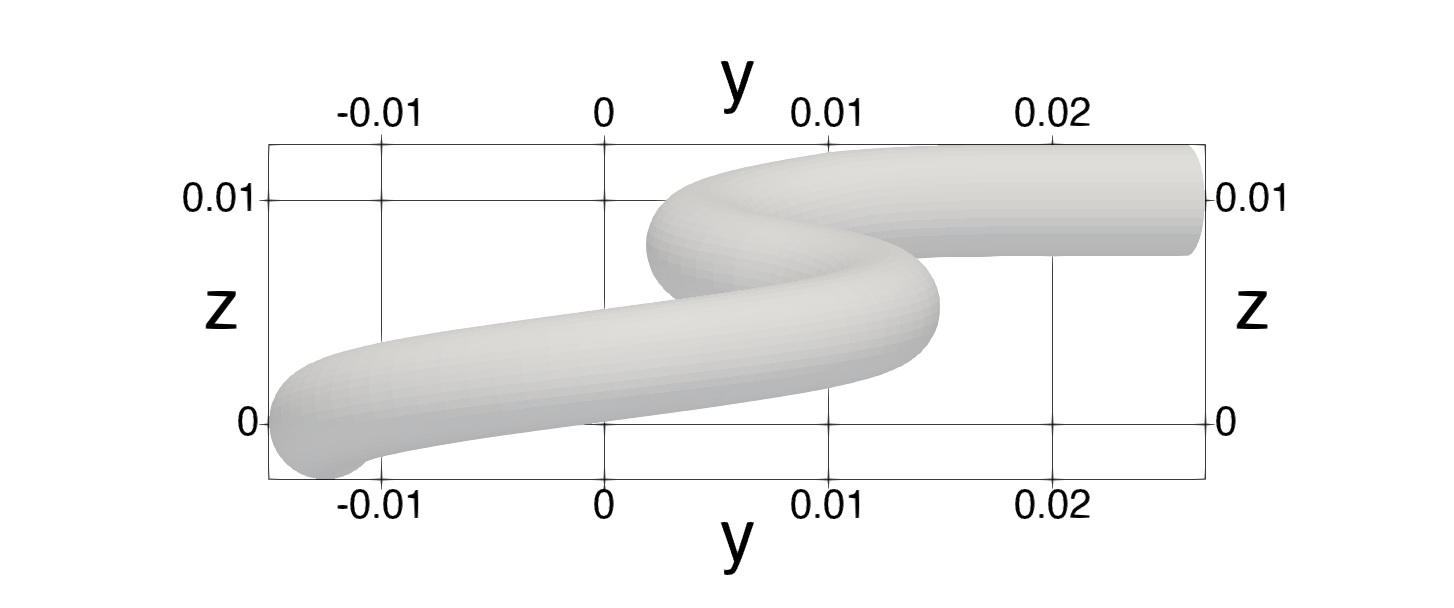}
     \end{subfigure}
     \caption{Optimal reactor geometry viewed from different perspectives. The optimal geometry contains an inversion.}
     \label{coil_res}
\end{figure}

The actual evaluated costs of each simulation generally fall beneath their predicted maximum cost, with the exception of a few simulations. 
The Gaussian process that models simulation cost can be validated as being accurate as a result. 
When the time remaining falls beneath the maximum predicted time to perform a standard iteration as well as a greedy final iteration, then the greedy solution is evaluated and the algorithm terminates.
The high-fidelity solution returned at the end of optimization is demonstrated in Figure \ref{coil_res}.
The optimal coil geometry has a pitch of 1.04cm, radius of 1.25cm, and an inversion that occurs 66\% along the coil length. 
The associated optimal operating conditions are pulsed-flow with an amplitude of 1mm, frequency of 2 Hz and a Reynolds number of 50. 

\begin{figure}[htb!]
    \centering
    \includegraphics[width=0.3\textwidth]{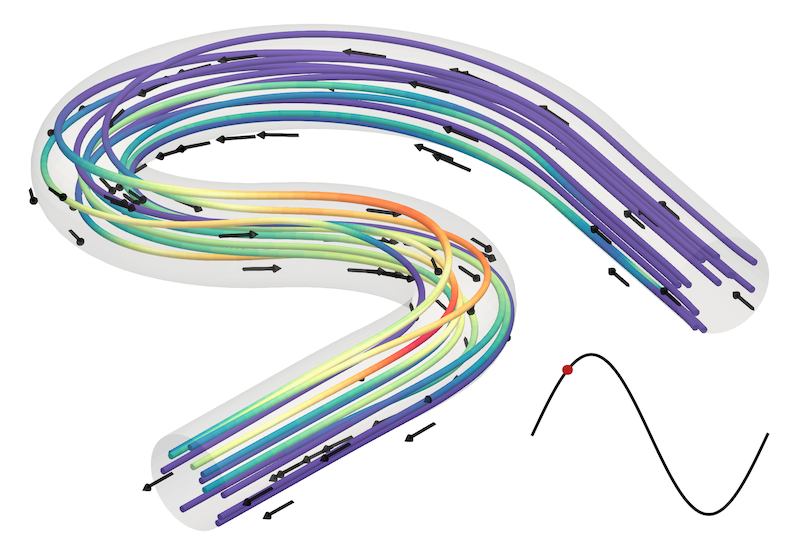}
    \includegraphics[width=0.3\textwidth]{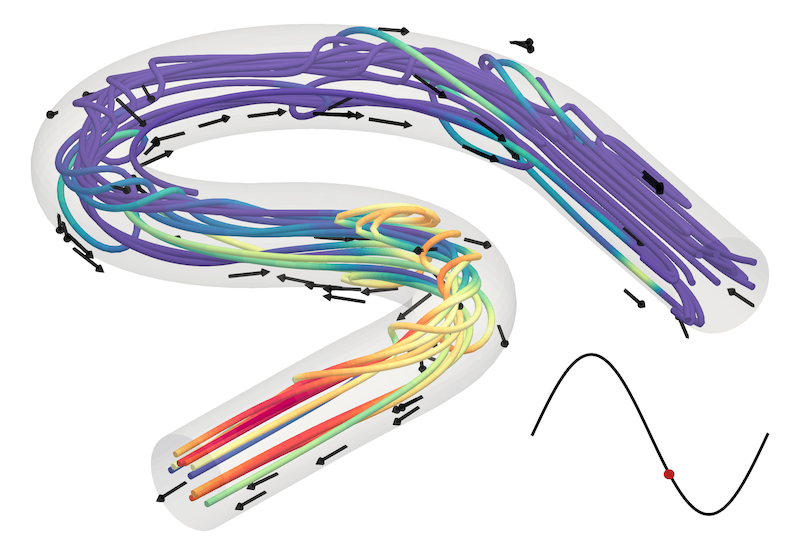}
    \includegraphics[width=0.3\textwidth]{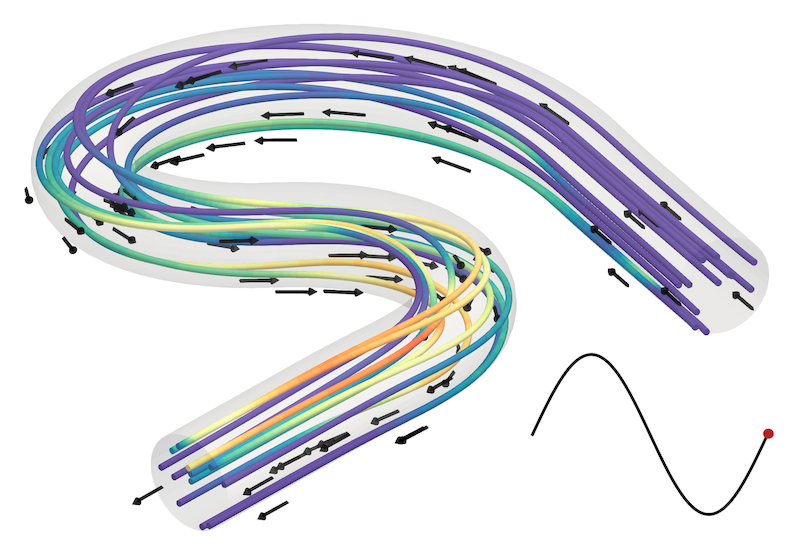}
    \caption{Streamlines indicating tracer concentration within the optimal reactor throughout a single oscillatory pulse, lasting 0.5 seconds. As the oscillatory velocity changes from positive to negative, Dean's vortices form providing mixing throughout the radial direction. The inversion within the coil also contributes to these vortices shifting across the radial direction, providing stronger mixing characteristics.}
    \label{fluids}
\end{figure}

Figure \ref{fluids} highlights the specific flow behavior within the optimal reactor throughout a single oscillatory cycle. 
Streamlines are coloured with tracer concentration to show the movement through the coil. 
For the positive part of the oscillatory cycle, the tracer moves in the forward direction in a streamlined manner towards the outlet. 
The important features start to occur in the negative part of the oscillation cycle where the secondary flow begins to emerge. 
Cross-sectional flow streamlines transitioning to Dean-type vortices due to the centrifugal forces at the coil turns are shown. 
These counter-rotating Dean vortices promote radial mixing of the tracer with the water medium and close to the walls of the computational domain; no tracer dead zones are left behind. 
Additionally, along with the reverse flow, the swirling motion developed during the negative oscillation cycle redirects the flow in the tangential direction which limits the axial dispersion of the tracer.  
This combination of promoting radial mixing and inhibiting axial mixing results in high plug flow performance. 
This flow cycle is periodic in nature for the amplitude of 1 mm and frequency of 2 Hz, and the simulation is continued until the tracer leaves the computational domain. 

In the optimal configuration shown in Figs. \ref{coil_res} and \ref{fluids}, the inversion just before the outlet helps in maintaining the Dean vortex structure until the flow leaves the domain. 
Whereas, in a standard configuration, the Dean vortices closer to the outlet could be disappearing gradually.  
In addition, the inversion helps in the release of a highly concentrated tracer that is trapped between the vortices to undergo further radial mixing. 
Therefore, enhanced radial mixing is achieved with this configuration due to the inversion, leading to an optimal plug flow performance.

\subsection{Solution Validation}

\begin{figure}[htb!]
    \centering
    \begin{subfigure}[b]{0.32\textwidth}
         \centering
         \includegraphics[width=0.95\textwidth]{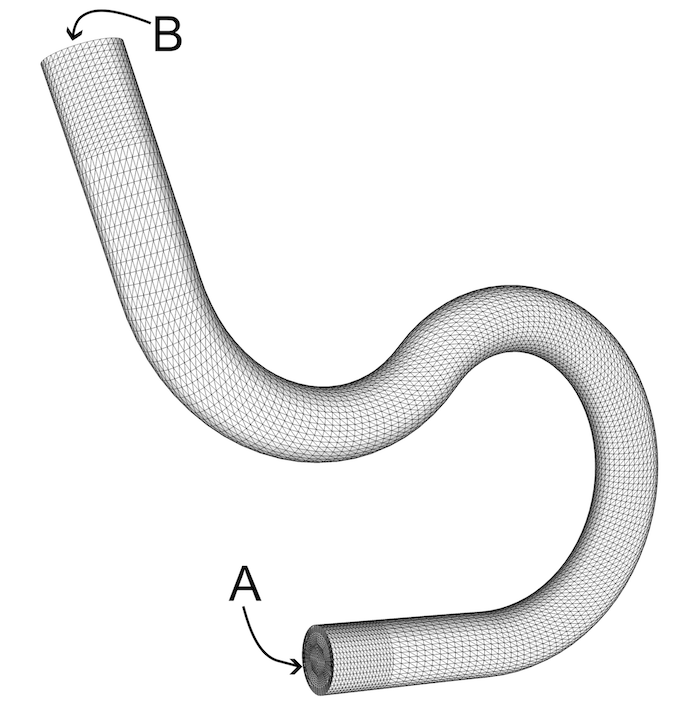}
         \caption{Raw STL geometry}
         \label{coil_stl}
     \end{subfigure}
    \begin{subfigure}[b]{0.32\textwidth}
         \centering
         \includegraphics[width=0.95\textwidth]{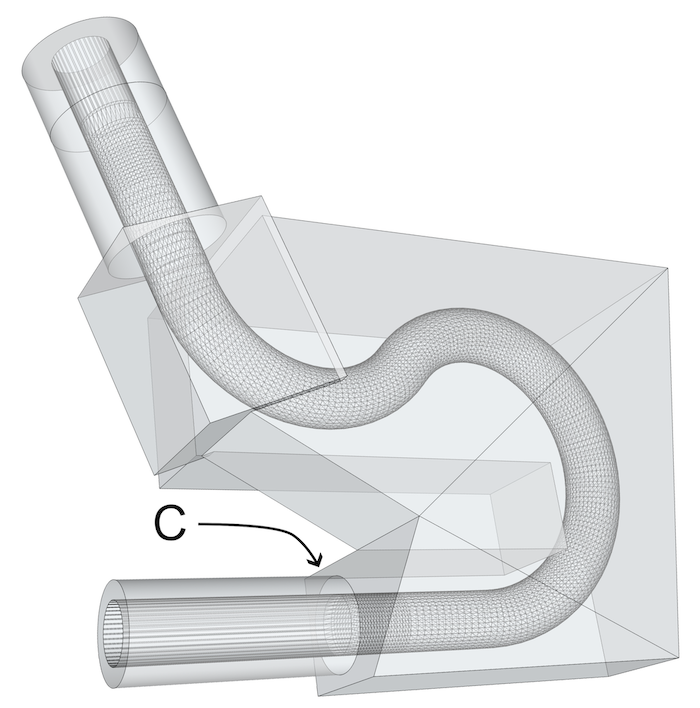}
        \caption{Modified STL geometry}
        \label{coil_b}
     \end{subfigure}
    \begin{subfigure}[b]{0.32\textwidth}
         \centering
         \includegraphics[width=0.95\textwidth]{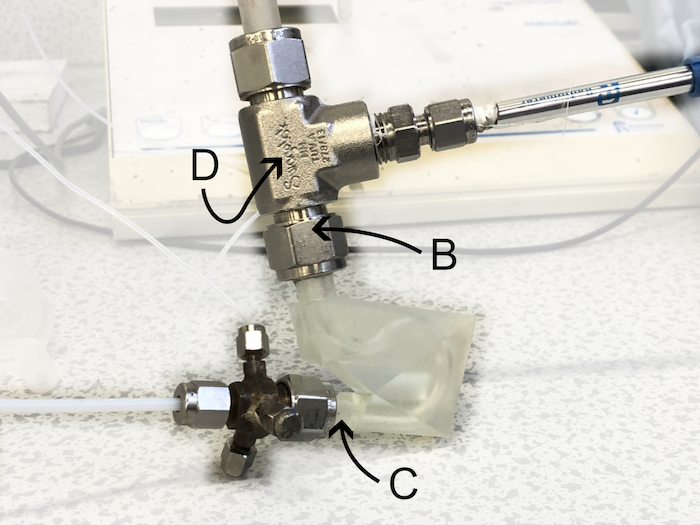}
        \caption{RTD measurements in the 3D-printed geometry.}
        \label{coil_exp}
     \end{subfigure}
     \caption{Experimental validation of optimal reactor configuration via additive manufacturing and operation under pulsed-flow operating conditions.}
     \label{coil_validation}
\end{figure}

The optimal reactor geometry was exported into an STL file format and modified into a 3D printable model by adding a bounding box to give the part volume. 
The straight sections at the inlet and outlet were given 8 mm and 10 mm OD tube fittings for connection to the experiment apparatus, and the bounding box was then trimmed to reduce the amount of resin needed for the print. The inlet region was also extended by 20 mm to provide additional development length to ensure the experiment matched as closely as possible the parabolic inlet velocity used in the simulations. 
The optimal geometry was printed using a FormLabs Form3+ using the Clear V4 resin using the default settings. 
Post-processing involved washing in IPA for 20 min, drying for 24 hours, and post-curing in a FormCure at 60°C for 30 min. 
Figure \ref{coil_validation} shows the raw STL file, modified geometry, and printed geometry. 

We implemented the same RTD method reported by \citet{McDonough2019a}.
In brief, RTDs were measured by injecting a 0.1 M KCl aqueous tracer solution into the geometry and measuring the conductivity over time at the outlet. 
The net flow of deionized water, oscillations, and tracer injection were controlled using three separate OEM syringe pumps (C3000, TriContinent) that were hydraulically linked to the reactor via PTFE tubing routed through a custom Swagelok piece (shown in Figure \ref{coil_exp}). 
To minimize the influence of poor mixing inside this Swagelok fitting, the tracer was injected at point C (Figure \ref{coil_validation}) after the additional 20 mm inlet straight section by routing the PTFE tubing through the fitting (see \citet{McDonough2019a} for further clarity).

\begin{figure}[htb!]
    \centering
    \includegraphics[width=0.5\textwidth]{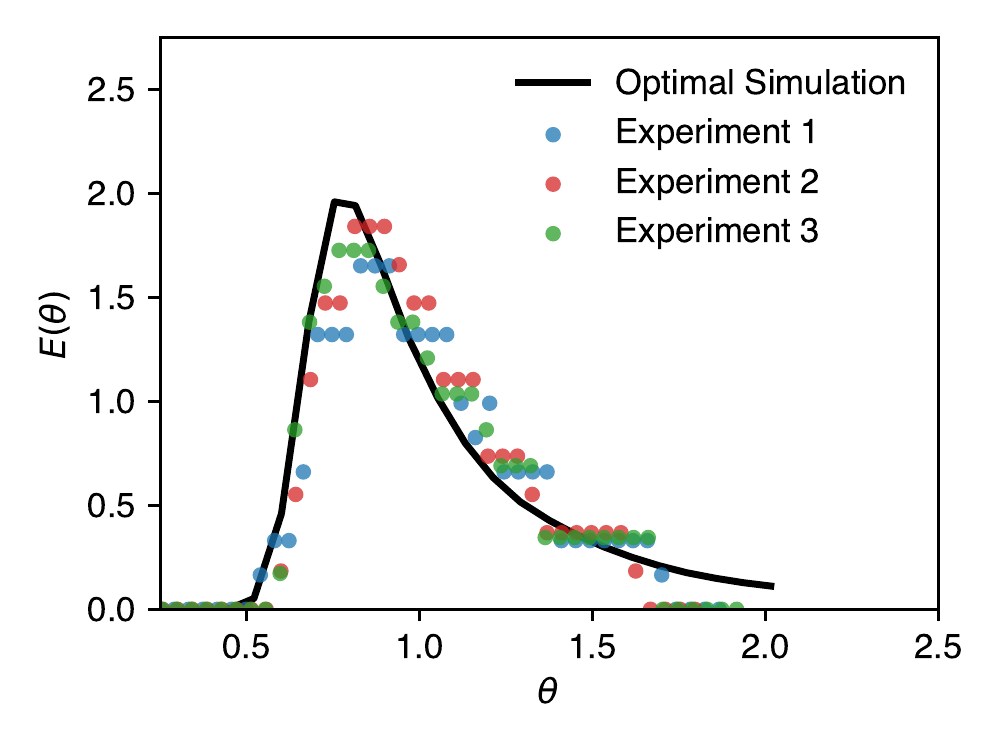}
    \caption{The residence-time distribution predicted via CFD simulation of the optimal, high-fidelity solution returned from the framework, alongside 3 sets of experimental data obtained via the 3D printed reactor.}
    \label{optimal_rtd}
\end{figure}

Figure \ref{optimal_rtd} demonstrates the predicted residence-time distribution alongside 3 sets of experimental data obtained from the 3D printed optimal reactor and operating conditions. We conjecture that the RTD method is another example of a lower-fidelity approximation to the desired mixing performance. 
First, the printed geometry is not perfectly smooth due to the well-known stair-stepping effect; this enhanced roughness (wall friction) will differ around the coil due to variations in the orientation. 
Second, the conductivity probe was located slightly downstream of the outlet face used in the simulations (point D in Figure \ref{coil_exp}). 
For reference, in the simulation the tracer was injected at point A and measured at point B (Figure \ref{coil_stl}). 
Nevertheless, the simulated optimal RTD closely matches the ‘expected’ distribution measured in the experiments across all three replicates, indicating good accuracy of the CFD simulations used to inform the optimization process. 
This provides confidence that the output of the optimization process is meaningful for a reactor geometry operating in a real-world setting.

\section{Conclusions and Future Work}\label{conc}

In this article, we have formulated the design of a helical-tube reactor as a multi-fidelity black-box optimization problem.
We have demonstrated a general framework that takes advantage of different quality simulations to enhance the optimization of reactor simulations via a multi-fidelity Bayesian optimization. 
We have validated our framework by applying it to optimize a helical-tube reactor geometry and operating conditions. 
By motivating our framework with the specific goal of returning a high-fidelity solution, we have derived a new criterion for monitoring the progress and dictating the termination of multi-fidelity Bayesian optimization. 
We show that this criterion enables a high-quality final solution to be returned before the computational budget is exhausted. 
The optimal reactor geometry and operating conditions is then 3D printed and the performance is experimentally validated.
Our approach is extensible to a large number of simulation-based design problems.
Future work will serve to apply the framework to investigate alternative reactor parameterizations and longer classes of reactors. 
In addition, different performance indicators will be optimized to investigate the differences between reactor parameterizations optimized for different quantities. 

\section*{Acknowledgements}
The authors would like to acknowledge the funding provided by the Engineering \& Physical Sciences Research Council, United Kingdom through the PREMIERE (EP/T000414/1). Tom Savage would like to acknowledge the support of the Imperial College President's scholarship, and Ilya Orson Sandoval for providing helpful discussion for this work.

\Urlmuskip=0mu plus 1mu\relax
\bibliography{main} 

\appendix
\section{Additive Manufacturing}
Additive manufacturing techniques have enabled the creation of reactors with less traditional, more complex geometries.
In order to recover these reactor geometries, highly-parameterized design formulations have to be created, resulting in large design spaces. 
To discover new reactor configurations with a view to manufacturing additively, these high dimensional spaces must be explored and optimized over.
\citet{ParraCabrera2018} reviews the use of additive manufacturing with an emphasis on the future of catalytic technologies. 
The authors also outline the important role optimization plays within additive manufacturing-based designs, highlighting the challenges faced with increasingly complex geometries. 
Helical-tube reactors \citep{McDonough2019a}, coil-in-coil reactors \citep{McDonough2019coilincoil} and toroidal reactors \citep{McDonough2020minitoroid} have all been proposed.
There is a need for robust tools to design and optimize complex geometries in order to support the next generation of additively manufactured mixers and reactors. 

\section{Pulsed-flow Helical-tube Reactors}
A promising class of reactor is the helical-tube reactor under pulsed-flow conditions.
Pulsed-flow conditions have been shown to induce Dean's vortices within the helical geometry. 
These vortices promote radial mixing within the tube and at the same time inhibit axial mixing, resulting in desirable plug-flow mixing characteristics \citep{McDonough2019a,McDonough2019coilincoil}. 
The pulsed-flow oscillation intensity has been suggested to cause vortex shedding or swirling flows which in turn impact plug flow performance.
\cite{McDonough2019a} confirms this by demonstrating plug-flow behavior switching ‘on’ and ‘off’ with changes in oscillation intensity.
Based on this, optimal operating conditions have been shown to occur at ratios of Deans to Reynolds numbers of between 2 and 8 and for Strohls numbers between 1 and 2 for the specific helical-tube reactor applied within the study.
However, the dependence on reactor geometry, as well as geometry optimization has not been performed.

\section{Affect of hyperparameters}

\begin{figure}[htb!]
    \centering
    \includegraphics[width=\textwidth]{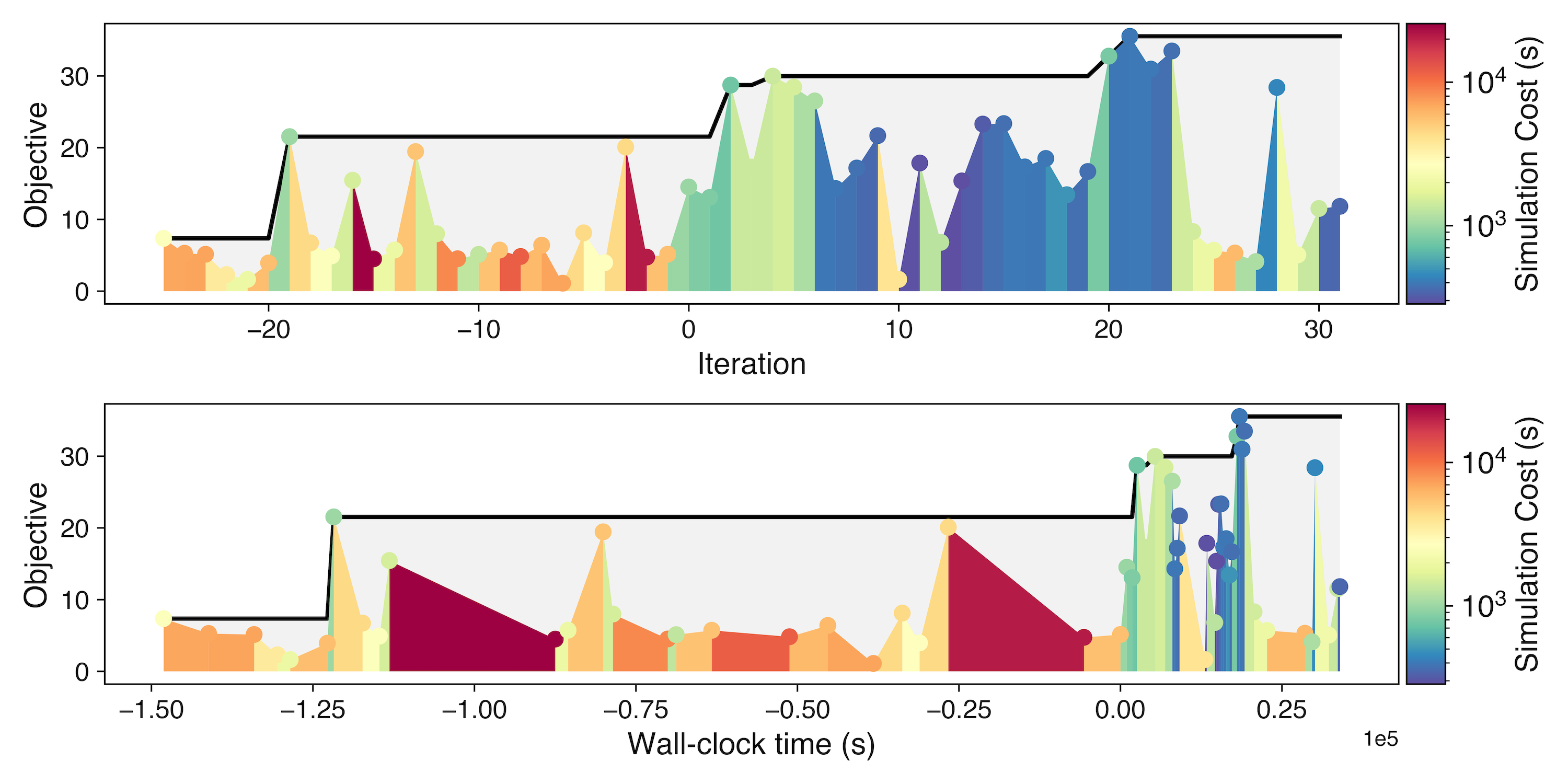}
    \caption{The number of equivalent tanks-in-series evaluated coloured by the respective cost of simulation. Optimization is performed with hyper-parameters $\gamma=1/2$ and $\beta=1/2$. The upper half of the figure shows these quantities against iteration and the lower half shows these quantities against wall-clock time, highlighting the importance of lower-cost simulations.}
    \label{fig:gamma12_beta12}
\end{figure}

The optimization procedure was repeated multiple times to investigate the affect of hyperparameters. 
These include $\beta$ controlling exploration tendencies, $\gamma$ controlling how cost is weighted, and $p_\lambda$ controlling how conservative the stopping criteria becomes. 
We use the same initial dataset generated to provide a more fair comparison between runs. 
Here we reduce the exploration parameter relative to the cost-adjusting parameter, with both being reduced relative to expected objective function, with $\gamma=1/2$, $\beta=1/2$.
Figure \ref{fig:gamma12_beta12} demonstrates the objective and cost of evaluated simulations throughout time.

\begin{figure}[htb!]
    \centering
    \includegraphics[width=\textwidth]{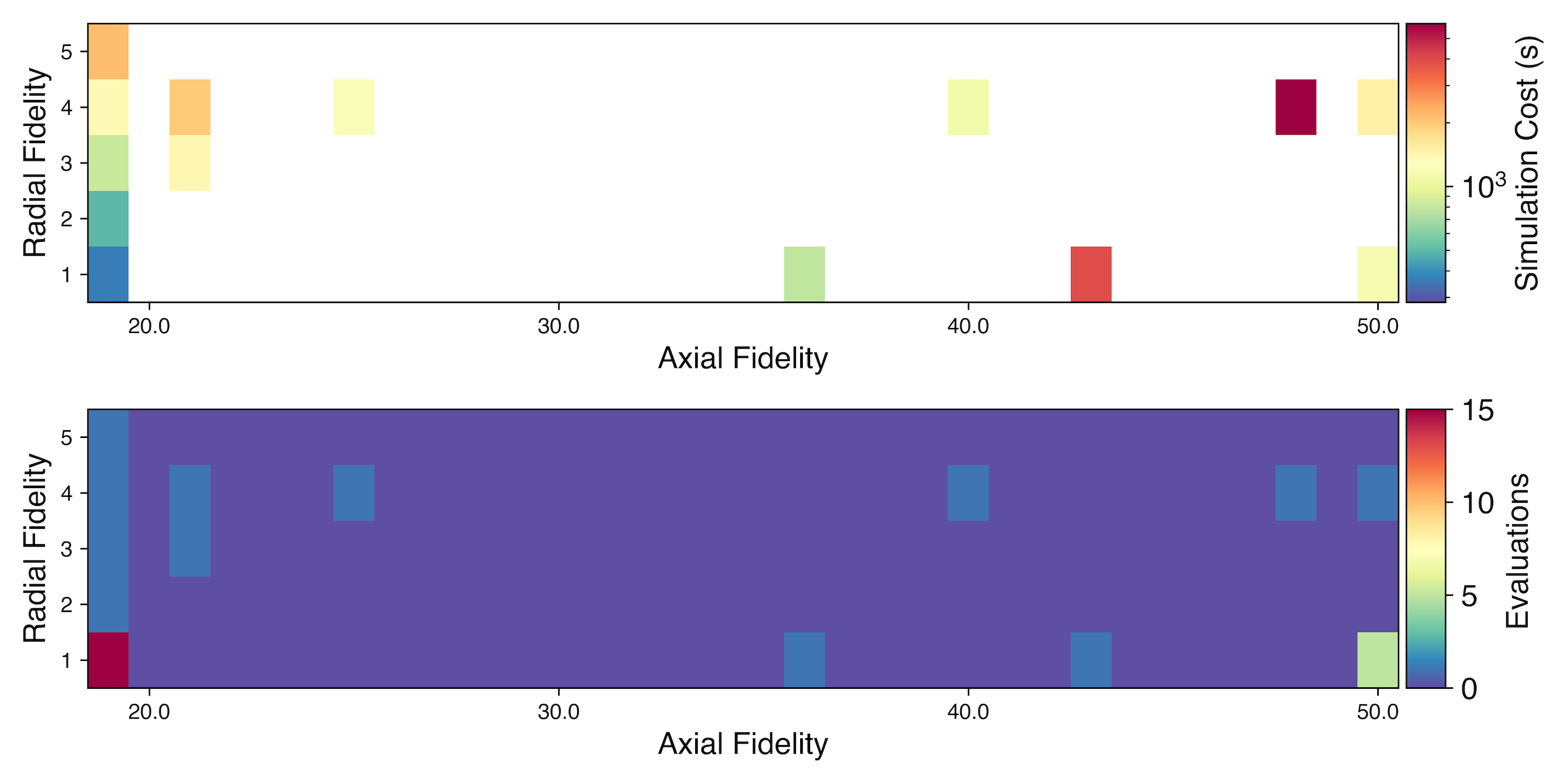}
    \caption{The fidelities selected throughout optimization within $\mathcal{Z}$. The upper plot demonstrates the average simulation cost at each fidelity. Optimization is performed with hyper-parameters $\gamma=1/2$ and $\beta=1/2$. }
    \label{fig:fidel1212}
\end{figure}

Figure \ref{fig:fidel1212} shows the fidelities selected throughout optimization, alongside average simulation cost, and number of evaluations.

\begin{figure}[htb!]
    \centering
    \includegraphics[width=\textwidth]{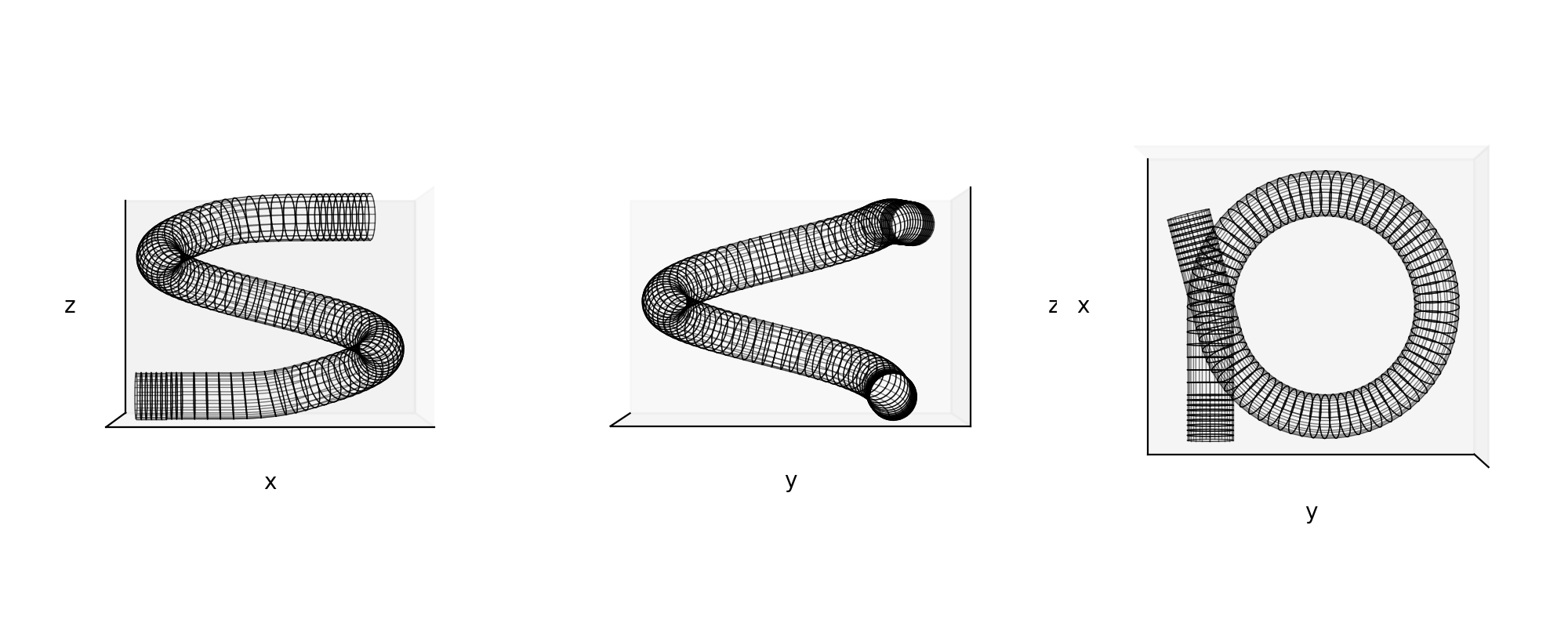}
    \caption{The optimal solution resulting from optimization with hyper-parameters $\gamma=1/2$ and $\beta=1/2$, the solution visually is similar to the solution presented in the main part of this article. The number of equivalent tanks-in-series of this configuration and operating conditions is 34.5.}
    \label{optimal_sol_two}
\end{figure}

Figure \ref{optimal_sol_two} demonstrates the geometry of the optimal reactor resulting from running the framework with this set of hyper-parameters.

\begin{figure}[htb!]
    \centering
    \includegraphics[width=\textwidth]{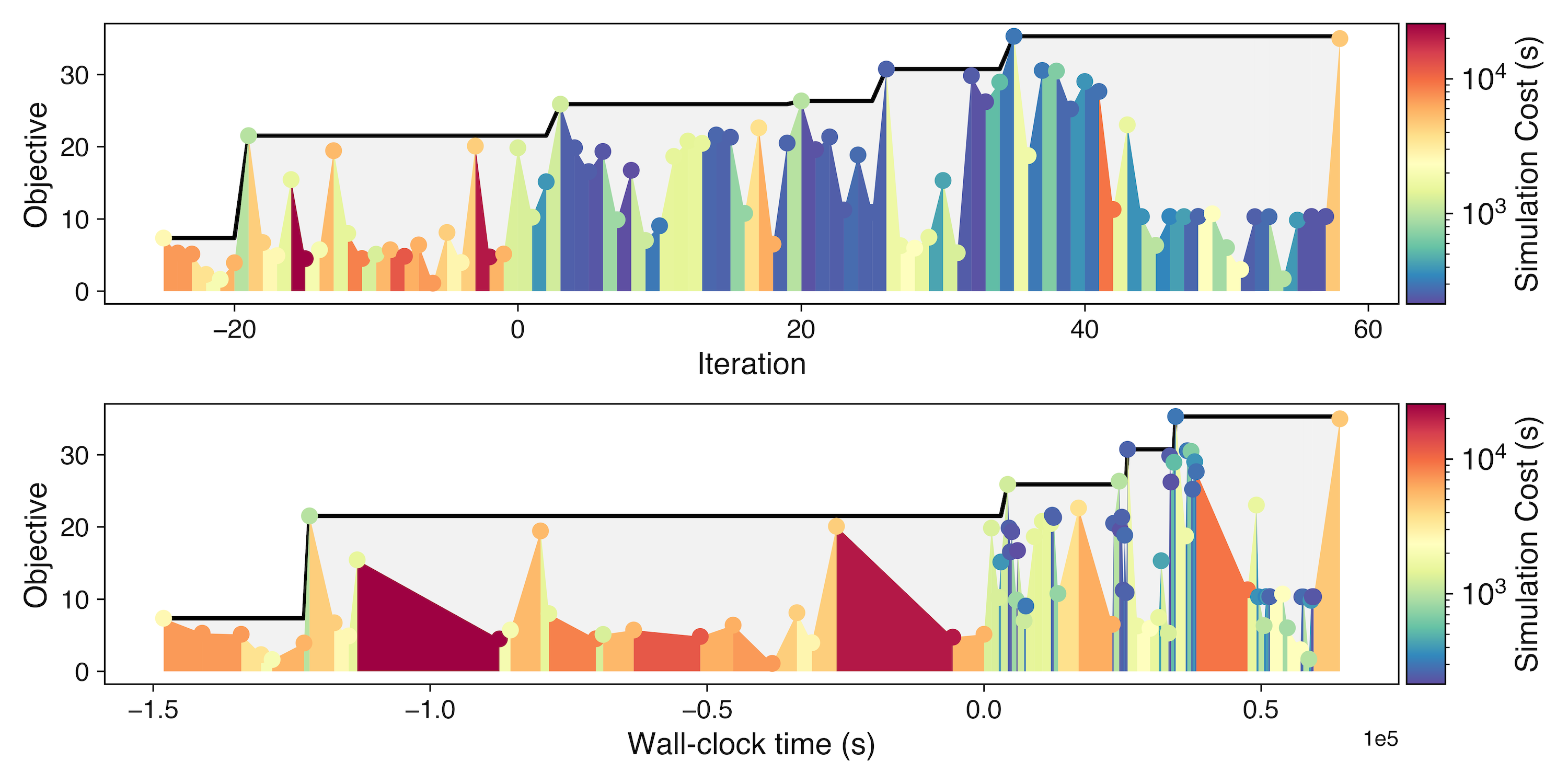}
    \caption{The number of equivalent tanks-in-series evaluated coloured by the respective cost of simulation. Optimization is performed with hyper-parameters $\gamma=1/2$ and $\beta=1$. The upper half of the figure shows these quantities against iteration and the lower half shows these quantities against wall-clock time, highlighting the importance of lower-cost simulations.}
    \label{fig:gamma12_beta1}
\end{figure}

Subsequently, we increase the exploration parameter relative to the cost-adjusting parameter and expected objective function, with $\gamma=1/2$, $\beta = 1$.
Figure \ref{fig:gamma12_beta1} demonstrates the objective and cost of evaluated simulations throughout time.

\begin{figure}[htb!]
    \centering
    \includegraphics[width=\textwidth]{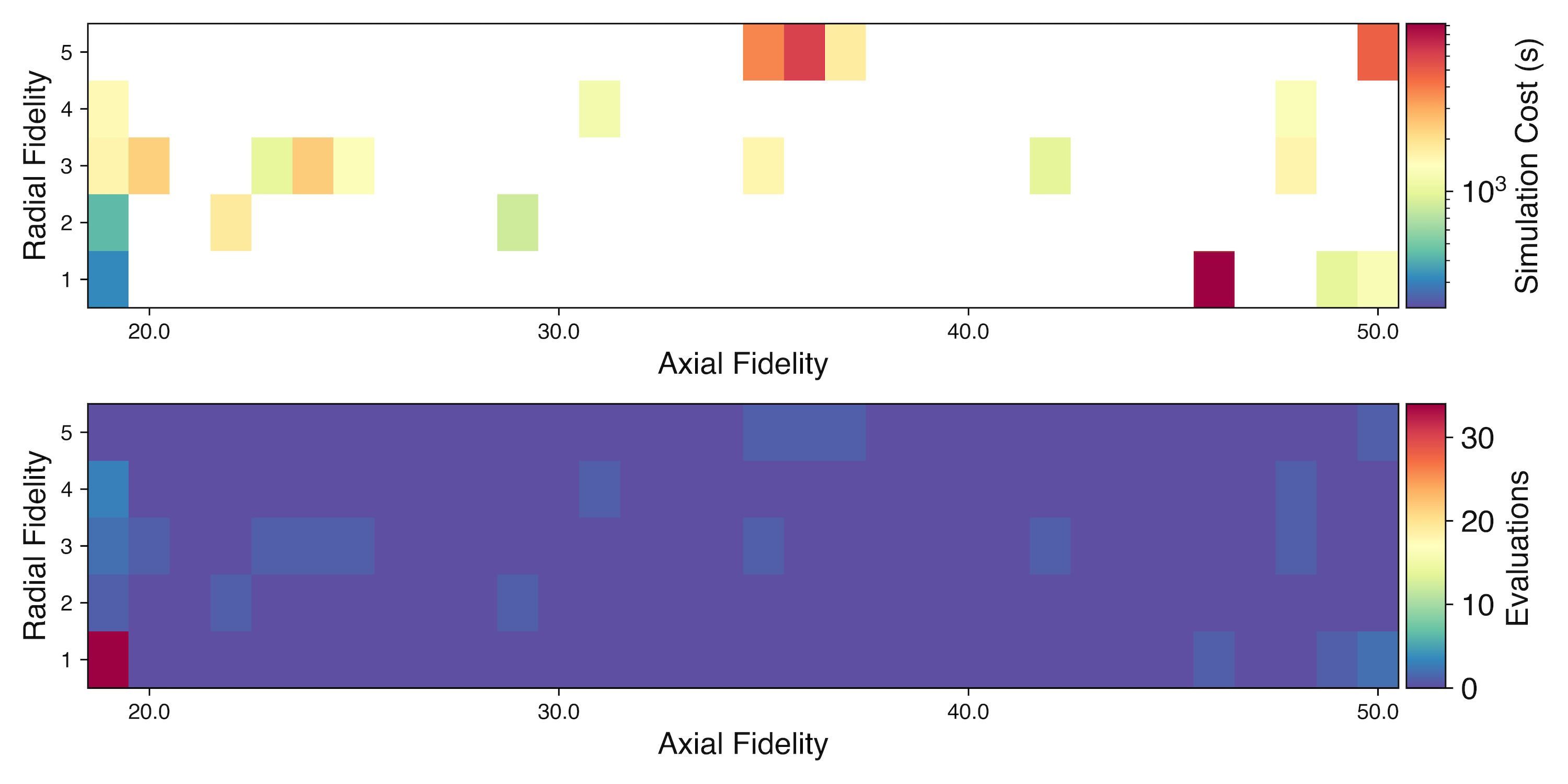}
    \caption{The fidelities selected throughout optimization within $\mathcal{Z}$. The upper plot demonstrates the average simulation cost at each fidelity. Optimization is performed with hyper-parameters $\gamma=1/2$ and $\beta=1$. }
    \label{fig:fidel12_1}
\end{figure}

Figure \ref{fig:fidel12_1} shows the fidelities selected throughout optimization, alongside average simulation cost, and number of evaluations.

\begin{figure}[htb!]
    \centering
    \includegraphics[width=\textwidth]{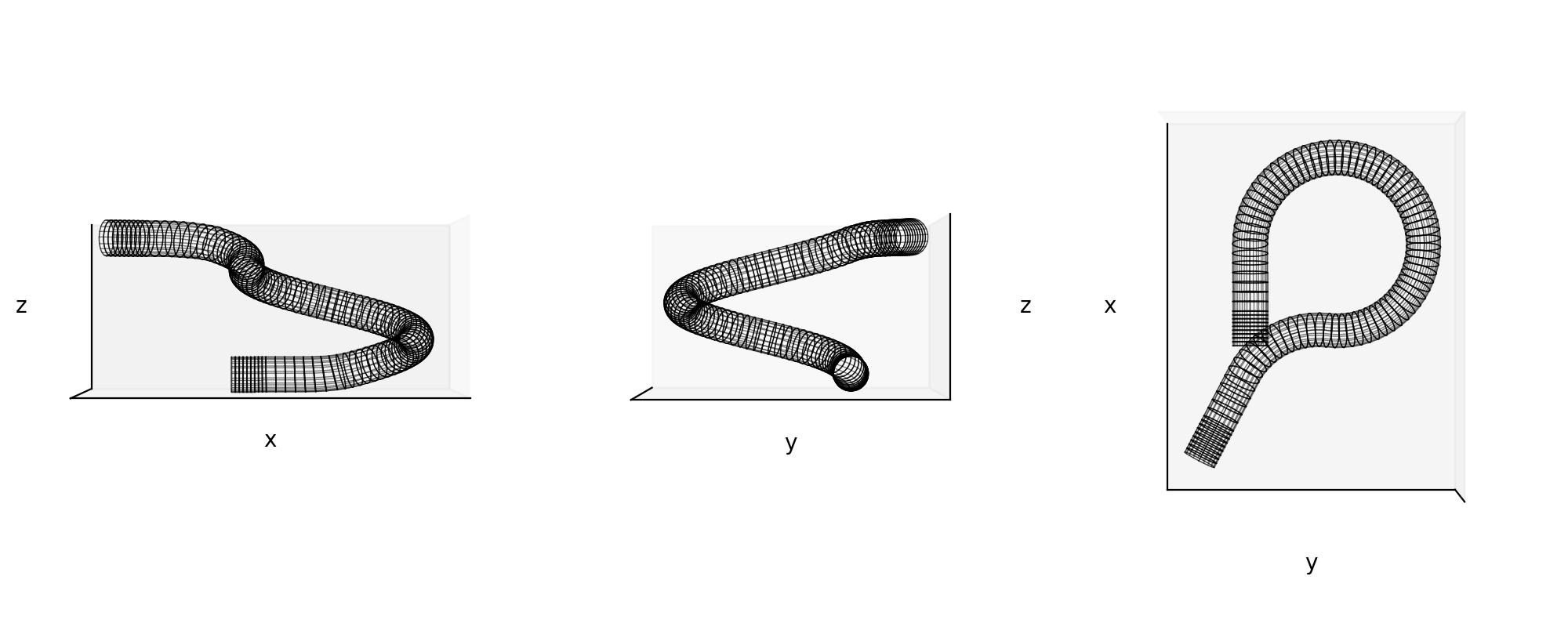}
    \caption{The optimal solution resulting from optimization with hyper-parameters $\gamma=1/2$ and $\beta=1$, the solution differs in that it does not contain an inversion, however the coil radius and operating conditions are consistent with the other optimal solutions. The number of equivalent tanks-in-series of this configuration and operating conditions is 26.8.}
    \label{optimal_sol_three}
\end{figure}

Figure \ref{optimal_sol_three} demonstrates the geometry of the optimal reactor resulting from running the framework with this set of hyper-parameters.

\end{document}